\documentclass[prd,showpacs,nofootinbib,tightenlines,preprintnumbers,superscriptaddress]{revtex4}
\usepackage{epsfig,graphics}
\usepackage{amssymb,amsmath,amsthm,graphicx,psfrag}

\newcommand{\Z}{{Z \!\!\! Z}}

\newcommand{\beqn}{\begin{eqnarray}}
\newcommand{\eeqn}{\end{eqnarray}}
\newcommand{\be}{\begin{equation}}
\newcommand{\ee}{\end{equation}}
\newcommand{\eq}[1]{(\ref{#1})}

\newcommand{\dd}{\mathrm{d}}

\newcommand{\dual}{\mbox{}^{\ast}}

\begin{document}

\preprint{HU-EP-04-22}
\preprint{ITEP-LAT/2004-09}
\preprint{LU-ITP 2004/015}

\title{
Phase Structure and Gauge Boson Propagator \\
in the radially active 3D compact Abelian Higgs Model}

\author{M.~N.~Chernodub}
\affiliation{ITEP, B.Cheremushkinskaja 25, Moscow, 117218, Russia}
\author{R.~Feldmann}
\affiliation{Institut f\"ur Theoretische Physik, Universit\"at
Leipzig, D-04109 Leipzig, Germany}
\author{E.-M.~Ilgenfritz}
\affiliation{Institut f\"ur Physik, Humboldt-Universit\"at zu Berlin,
Newtonstr. 15, D-12489 Berlin, Germany}
\author{A.~Schiller}
\affiliation{Institut f\"ur Theoretische Physik, Universit\"at
Leipzig, D-04109 Leipzig, Germany}


\begin{abstract}
Unfreezing the radial degree of freedom, we study the three-dimensional Abelian
Higgs model with compact gauge field and fundamentally charged matter.
For small quartic Higgs self coupling and finite gauge coupling the model
possesses a first order transition from the confined/symmetric phase to the
deconfined/Higgs phase separated at some hopping parameter $\kappa_c$.
Latent heat and surface tension are obtained in the first order regime.
At larger quartic coupling the first order transition ceases to exist, and
the behavior becomes similar to that known from the London limit.
These observations are complemented by a study of the photon propagator
in Landau gauge in the two different regimes.
The problems afflicting the gauge fixing procedure are carefully investigated.
We propose an improved gauge fixing algorithm which uses a finite subgroup
in a preselection/preconditioning stage.
The computational gain in the expensive confinement region is a speed-up
factor around 10.
The propagator in momentum space has a non-zero anomalous dimension
in the confined phase whereas it vanishes in the Higgs phase.
As far as the gauge boson propagator is concerned, we find that the radially
active Higgs field provides qualitatively
no new effect compared to the radially frozen Higgs field studied before.
\end{abstract}


\maketitle


\section{Introduction}
The lattice Abelian Higgs model with compact gauge fields (cAHM) has a wide
variety of properties which makes it interesting from both the high energy
physics~\cite{Fradkin,EinhornSavit} and
condensed matter physics~\cite{AnomalousMatter} point of view.
The compactness of the gauge field gives rise to the presence of monopoles
which are instanton-like excitations (topological defects) in
three space-time dimensions. Being in the plasma state, monopoles and antimonopoles
guarantee linear confinement of electrically charged test particles~\cite{Polyakov}.
The topological defects are forming an oppositely charged double sheet along the
minimal surface spanned by a Wilson loop. Due to screening, the free energy
of this double sheet is proportional to the area of the minimal surface.
As a result, electrically charged particles experience linear confinement.

In the case of the cAHM, the plasma state of the monopoles is realized, similar
to pure cQED, at small values of the hopping parameter (coupling between matter
and gauge fields) corresponding to the confined phase of cAHM.
With respect to the Higgs degrees of freedom that phase is called symmetric.
As the hopping
parameter increases, the system enters the Higgs region, where {\it e.g.} the
linear potential between test charges is suppressed.
Correspondingly, in this phase monopoles and antimonopoles are bound into
magnetically neutral dipoles, which provide at best only short-ranged
interactions between the charged test particles. In other words, the presence
of dynamical matter fields with charges in the fundamental representation of the
gauge group leads to the effect of string breaking which
results in a flattening of the potential at large charge separations.
With respect to the gauge degrees of freedom the Higgs region
is the deconfined region. The formation of magnetic dipoles
in relation to the phenomenon of string breaking were studied both analytically
in Ref.~\cite{AnomalousMatter} and numerically in Ref.~\cite{Chernodub:2002ym}.
Note that the origin of monopole binding in the zero temperature case of cAHM$_3$
is physically different from the case of monopole binding at the finite
temperature phase transition occurring in compact QED$_{2+1}$ without a dynamical
Higgs field in the fundamental representation~\cite{Agasian:2001an,Binding,CIS12}.

The effect of finite-temperature deconfinement on the photon propagator in
($2+1$)-dimensional cQED~\cite{CIS3,Chernodub:2002gp,Chernodub:2003bb} and in
zero-temperature cAHM in the London limit are rather
similar~\cite{Chernodub:2002ym}. The propagator is described by a Debye mass
and by an anomalous dimension which both vanish at the deconfinement transition
regardless on its nature, which can be caused
by finite temperature or by the presence of the matter fields. In both cases
the transition is attributed to pairing of magnetic monopoles. The mass parameter
of the propagator behaves differently in these cases since in the deconfined
phase of the cAHM the gauge boson acquires a mass due to the Higgs mechanism
whereas in the case of cQED this mechanism is of course absent.

The qualitative similarity of the form of the gauge boson propagators in
cQED and in the cAHM, so far known in the London limit, immediately raises the
question about the role of the Higgs field in the emergence of the anomalous
dimension. Indeed, in the London limit of the cAHM the mass of the Higgs is
infinite and the only active ingredient of the Higgs field is its compact phase.
Away from the London limit the Higgs mass stays finite and the radial component 
of the Higgs field gains influence. In this paper we concentrate on the
role of the radially active Higgs field on the properties of the gauge boson
propagator in conjunction with the phase structure of the model.

The phase diagram of the cAHM in three dimension has been extensively studied
both perturbatively (analytically) and numerically in the
literature~\cite{ref:phase,ref:phase:prop1,ref:phase:prop2}.
Already in the early studies~\cite{ref:phase} the changing nature of the phase 
transition due to radially active Higgs field has been addressed.
The authors of Refs.~\cite{ref:phase:prop1,ref:phase:prop2} have been concentrated
on the continuum limit (studying large $\beta$) measuring among others
the gauge boson mass from plaquette-plaquette correlators. As in our previous work
we are studying here the propagator in momentum space which allows to establish
easily whether a non-vanishing anomalous dimension exists.

The structure of the paper is as follows. Section~\ref{sec:model} is devoted to
the description of the model and to a sketch of its phase structure at two
selected values of the coupling describing the quartic self-interaction of
the Higgs field.
The gauge fixing procedure and its inherent problems are
reviewed in Section~\ref{sec:gauge} where also a preconditioning method using
a finite subgroup of the gauge group is proposed.
Near the transition, the changing form of the photon propagator is analysed
in Section~\ref{sec:results}. Our conclusions are given in
Section~\ref{sec:conclusions}.

\section{The Model and its Phase Structure}
\label{sec:model}

\subsection{Some properties of the model and its limiting cases}

We consider the three-dimensional Abelian Higgs model
with compact gauge fields $\theta_{x,\mu} \in [-\pi,\pi)$ living on
links $l={x,\mu}$ and with fundamentally charged ($Q=1$) Higgs fields
$\Phi_x \in \mathbb{C}$ on sites $x$ (a vector with integer Cartesian coordinates).

The Higgs field is written in the form
\be
\Phi_x= \rho_x {\rm e}^{i \varphi_x}
\ee
where $\rho_x$ is its radial part and $\varphi_x$ its phase.
The model is defined by the action
\beqn
  S = - \beta \sum_P \cos\theta_P
            - \kappa \sum_{x,\mu} \rho_x \rho_{x+\hat{\mu}}
        \cos(- \varphi_x +
        \theta_{x,\mu} + \varphi_{x+\hat{\mu}})
        +\sum_x (\rho_x^2 + \lambda (\rho_x^2-1)^2) \,,
  \label{eq:action}
\eeqn
where $\theta_P$ is the plaquette
angle representing the curl of the link field $\theta_{x,\mu}$ and
$\hat{\mu}$ is the unit vector in $\mu$ direction.
$\beta$ is proportional to the inverse gauge coupling squared, 
$\beta=1/(a~g^2)$,
$\kappa$ is the hopping parameter and $\lambda$ the quartic
Higgs self coupling. The so called London limit of that model has frozen 
radial Higgs length corresponding to $\lambda \to \infty$.

In that limit, at zero value of the hopping parameter $\kappa$, the model
(\ref{eq:action}) reduces to the pure compact Abelian gauge theory
which is known to be confining at any coupling $\beta$ due to the
presence of the monopole plasma~\cite{Polyakov}. We call the low--$\kappa$
region of the phase diagram the ''confined region''. At large values of
$\kappa$ (also called the ''Higgs region'') the monopoles should disappear
because the requirement $\kappa \rightarrow\infty$ leads to the constraint
$- \varphi_x +   \theta_{x,\mu} + \varphi_{x+\hat{\mu}}=2\pi{}l_{x,\mu}$,
with, in general, $l_{x,\mu}\in \Z$. However, in the unitary gauge,
$\varphi_x =0$, the above constraint along with the compactness condition
for the gauge field, $\theta_{x,\mu} \in [-\pi,\pi)$, gives
$\theta_{x,\mu}= 2\pi{}l_{x,\mu} = 0$, which, in turn, indicates the
triviality of the model at large values of $\kappa$ in the London limit.

At large $\beta$ the model with $\lambda \to \infty $ reduces to the three
dimensional $XY$ model which is known to have a second order phase transition at
$\kappa^{XY}_c \approx 0.453$~\cite{XYphase} between a symmetric and a Higgs phase.
Indeed, in the limit $\beta \to \infty$ we get the constraint for the plaquette
variable\footnote{Here and below we often use the differential form
notation on the lattice $(a,b) = \sum_l a_l\, b_l$,
${||\theta||}^2 = (\theta,\theta)$, the operators
$\dd$ and $\delta$, respectively, the lattice curl and divergence.
The Laplacian is denoted as $\Delta = \delta \dd + \dd \delta$.
}, $\theta_P \equiv {[\dd \theta]}_P= 2 \pi n_P$, with $n_P \in \Z$. The constraint
implies $\dd n = 0$ (due to the nilpotency $\dd^2 = 0$),
which, in turn, gives $n=\dd m$, where $m\in \Z$ is a link variable. Thus, the
constraint reduces to the equation $\dd (\theta - 2 \pi m) = 0$ which has a
general solution of the form (in usual notations)
$\theta_{x,\mu} = - \chi_{x+\hat\mu} + \chi_{x} + 2 \pi l_{x,\mu}$,
where $\chi_{x} \in [-\pi,\pi)$ is a compact scalar field. The integer--valued
vector field, $l_{x,\mu} \in \Z$, is chosen in such a way
that $\theta_{x,\mu} \in [-\pi,\pi)$. Fixing the unitary gauge we obtain that
in the London limit and at large $\beta$ the action~\eq{eq:action} reduces to 
the action of the $XY$--model,
$S^{XY} = - \kappa \sum_{x,\mu} \cos(- \chi_x + \chi_{x+\hat{\mu}})$,
where the scalar angle $\chi$ plays the role of the spin field in that model.

{}From our previous studies in the London limit at
$\beta=2.0$~\cite{Chernodub:2002ym,Chernodub:2002en} we know
that the transition from the confined to the Higgs region is signalled by a
rapid drop of the monopole density.
In addition, a string breaking phenomenon affecting fundamental test charges 
(being in the same representation as the fundamentally charged matter
fields) has been observed which are bound by a linear (string) potential in 
the confined region.
While the drop of the monopole density signals onset of deconfinement,
the anomalous dimension of the photon propagator turning to zero at the same
critical $\kappa_c$ shows the transition from the symmetric to the Higgs region,
whereas the observed non-zero photon mass at larger $\kappa$ is simply due to
the onset of normal mass generation by the Higgs effect.

However, we could not confirm that this crossover is accompanied by an ordinary
phase transition. It cannot be excluded that the transition is second order with
a very small negative critical exponent or possibly of Kosterlitz-Thouless
type.

Including a fluctuating radial degree of freedom $\rho_x$ (at small $\lambda$)
changes the situation strongly. For fixed $\lambda$ a first order phase transition 
is known to exist~\cite{ref:phase:prop2}
from the symmetric to the Higgs phase in the $\beta$--$\kappa$ plane
accompanied by the monopole density approaching zero, similar to our previous
studies of the $3D$ SU(2)-Higgs model~{\cite{Ilgenfritz:1995sh,Gurtler:1996wx}}.
With increasing quartic Higgs self coupling the first order transition
becomes weaker and ends at a certain critical $\lambda_c$~\cite{Gurtler:1997hr}.
Above that critical self coupling the phase structure of the compact Abelian 
Higgs model has been investigated in~\cite{ref:phase:prop1} for relatively 
large $\beta$ with practically vanishing monopole density.
In a very recent study~\cite{wenzel} the location of $\lambda_c$  has been
established for a smaller (inverse) gauge coupling $\beta=1.1$ as
\be
0.030 < \lambda_c  < 0.032 \,.
\ee

\subsection{Two regimes of Higgs self coupling}

In the present work we consider the gauge coupling $\beta=2.0$ corresponding
to a smaller lattice spacing $a$ than in Ref.~\cite{wenzel}. 
For this gauge coupling the endpoint of the transition is found to be~\cite{future}
\be
  \lambda_c \approx 0.009(1) \,.
\ee
More details of this phase transition study will be presented
elsewhere. We restricted the Higgs self coupling to two
values $\lambda=0.005$ (representing the first order transition regime)
and $\lambda=0.02$ (representing the continuous transition region).

In our simulations one complete Monte Carlo update consists of a Higgs and a
gauge field update in alternating order. For the gauge fields a usual Metropolis
algorithm has been applied. The algorithm was adjusted (during the thermalization)
to ensure an acceptance rate of about 50 per cent.
Representing the Higgs field as real two-component vector a heat-bath algorithm
similar to the method described in~\cite{Bunk:xs} has been used. Additional
overrelaxation steps have been applied to the Higgs field. Since the autocorrelation
changes marginally with the number of reflection steps not more than two reflections per heatbath update have been used.

In order to locate the phase transition and to quantify its strength we consider
for the present purposes only the gauge invariant average Higgs modulus squared
\be
\rho^2= \frac{1}{L^3} \sum_x  \rho_x^2 \,
\ee
($L$ is the linear extent of the lattice)
and evaluate its vacuum expectation value $\langle \rho^2 \rangle$.
For the Higgs modulus squared one can consider the susceptibility
\be
C_{\rho^2}= L^3\, \left\{\langle (\rho^2)^2 \rangle -\langle \rho^2 \rangle^2\right\}
\ee
and the Binder fourth cumulant as function of $\kappa$ (at fixed $\beta$ and
$\lambda$) near criticality for different lattice volumes.
Widely used multihistogram reweighting methods have also been applied in our work
in order to determine the pseudocritical hopping parameter $\kappa_c$.

\begin{figure*}[!htb]
\begin{center}
\begin{tabular}{cc}
\includegraphics[scale=0.35,clip=true]{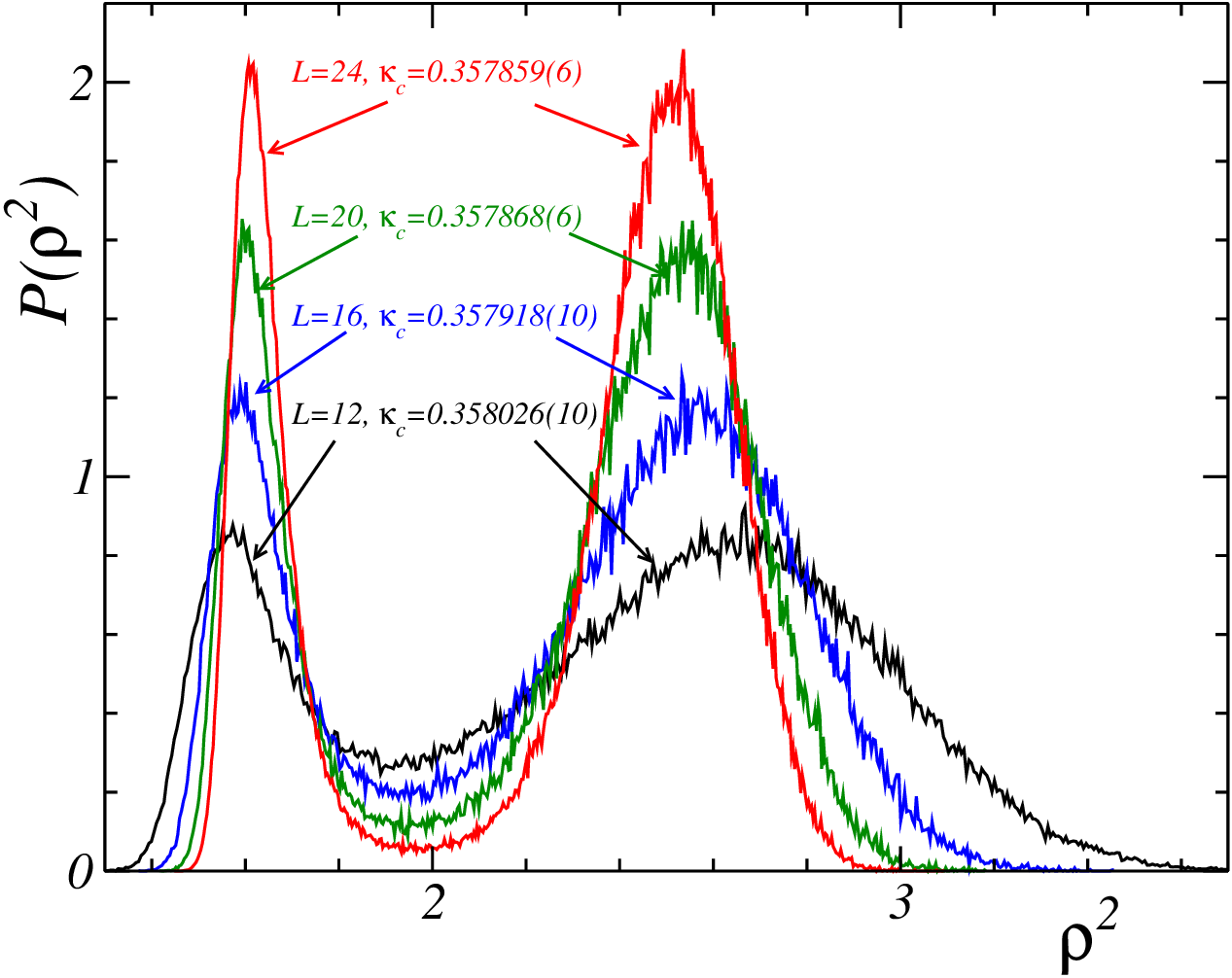} &
\includegraphics[scale=0.35,clip=true]{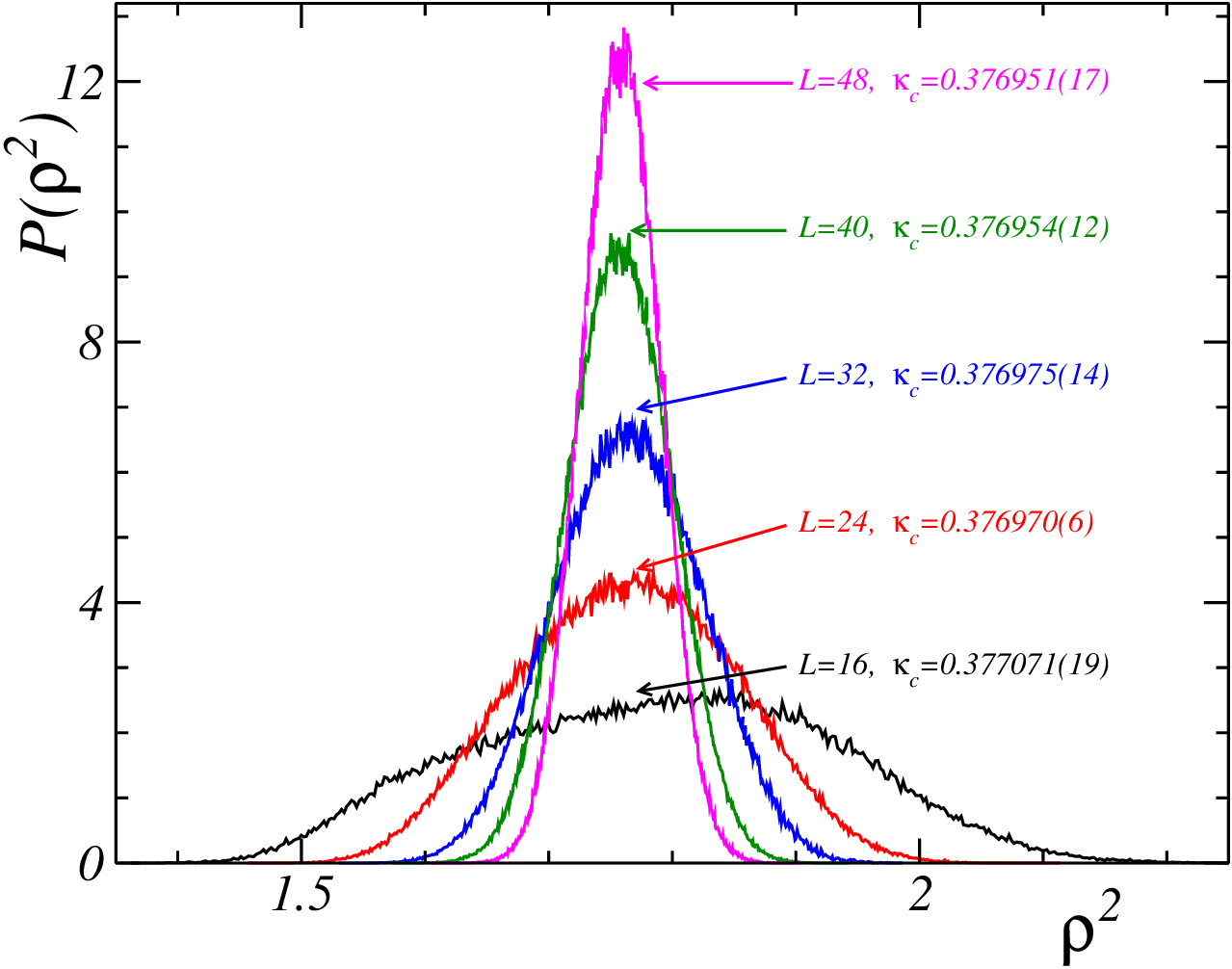}
\end{tabular}
\end{center}
\caption{$\rho^2$ histograms at pseudocriticality:
(a) for $\lambda=0.005$ with $\kappa_c$ fixed according to the equal height
criterion and
(b) for $\lambda=0.02$ according to the maximal susceptibility criterion.}
\label{fig:1}
\end{figure*}

As one example we show here in Fig.~\ref{fig:1} for the two $\lambda$ values
the histograms $P(\rho^2)$ (normalized to unit area) from all data at different
$\kappa$ values near $\kappa_c$ and lattice sizes $L^3$ reweighted to the
respective pseudocritical hopping parameter $\kappa_c$. In case (a) $\kappa_c$
for the lower
$\lambda$ has been fixed according to the equal height condition
applied to the two-state histogram.
In case (b) for the larger
$\lambda$ value $\kappa_c$ has been determined according
to the condition that the susceptibility $C_{\rho}$ takes its maximum.
One observes that in Fig.~\ref{fig:1} (a) the two peaks become more pronounced
with increasing volume whereas the distance between the maxima only marginally
decreases. The tunnelling frequency between the two coexisting phases in a finite
system (obtained from the autocorrelation function of $\rho^2$)
rapidly decreases with increasing volume.
At a lattice volume $40^3$ (the one used for the propagator measurements
to be discussed later) no tunnelling has been observed. This has allowed us
to cleanly study and contrast the respective propagators of the competing phases.

The first order transition is characterized by a finite jump
$\Delta \langle \rho^2 \rangle$ of the average Higgs modulus squared,
a two-state signal at pseudocriticality, surviving in the infinite volume limit.
This jump is related to the latent heat of the first order transition
\be
\frac{L_{\rm heat}}{g_3^2}=\frac{1}{2} \kappa \beta \, \Delta \langle \rho^2 \rangle
\ee
with $g_3^2=1/(a \beta)$,
$a$ is the lattice spacing. In the symmetric phase $\langle \rho^2 \rangle $ is 
nearly constant as a function of $\kappa$.
In contrast to this, on the Higgs side of the transition
$\langle \rho^2 \rangle $ increases approximately linear with $\kappa$.

Fig.~\ref{fig:1} (a) clearly demonstrates that the transition is first order
at $\lambda=0.005$. We also have observed that the maximum of the susceptibility
(not shown here) increases linearly with the volume.
Following Ref.~\cite{ref:phase:prop2} we can estimate the dimensionless
surface tension from the shape of the equal height histogram:
\be
\sigma= \frac{ \beta^2}{2 L^2 } \ln \frac{P_{\rm max}(\rho^2)}
                                         {P_{\rm min}(\rho^2)} \,.
\ee
Here $2 L^2$ is the dimensionless area of two minimal surfaces splitting the
periodic lattice into two regions filled with the two pure phases being in
coexistence;
$P_{\rm max}(\rho^2)$ and $P_{\rm min}(\rho^2)$ are the maximal/minimal
value of the histogram, respectively. An extrapolation to the infinite volume
limit (taking into account the leading $1/V$ corrections) leads
to the following estimates of the latent heat and of the dimensionless 
surface tension
\be
\frac{L_{\rm heat}}{g_3^2}=  0.320(2) \,, \quad \sigma= 0.0121(3) \,.
\ee

Fig.~\ref{fig:1} (b) shows immediately that at $\lambda=0.02$ a first order
transition does not exist. As indicated in the figure the histogram merges
into definitely one maximum at larger lattice volume. Therefore there is no
non-vanishing latent heat. The width of the histogram
reduces with increasing volume. Within errors, the susceptibility
for the largest studied volumes (again not shown) does not
change as function of the volume.
Fitting the susceptibility at the maximum for lattices from $L=32$ to $L=48$ 
by the function $C_{\rho^2} = A L^{\gamma}$  gives $\gamma=-0.014(40)$ with
$\chi^2/d.o.f. = 0.9$.
Including a subleading correction we obtain $\gamma=-0.012(27)$ for lattices 
from $L=16$ to $L=48$.
This excludes a second order transition with a critical exponent differing
significantly from zero. The transition is similar as the one found in
the London limit\cite{Chernodub:2002ym}.

\subsection{The monopole density near criticality and autocorrelation estimates}

In addition to $\rho^2$ we also consider the monopole degrees of freedom
constructed out of the gauge fields. Using the differential form notation
we consider the monopole charges as the 0-form $\dual j \in \Z$
defined on the sites dual to the lattice cubes $c$~\cite{DGT} where it is
evaluated as follows:
\beqn
  j = \frac{1}{2\pi} \sum\limits_{P \in \partial c} {(-1)}^P \,
  {[\theta_P]}_{\mathrm{mod} \, 2 \pi} \,.
\label{jc}
\eeqn
The factor ${(-1)}^P$ takes the plaquette orientations
relative to the (outer) boundary of the cube into account,
and the notation $[\cdots]$ means taking the integer part modulo $2\pi$.
The 2-form
\be
p[j] = {[\dd \theta]}_{\mathrm{mod} \, 2 \pi} 
\label{Diracstrings}
\ee
corresponds to the Dirac strings living on the links of the dual lattice,
which are dual to the plaquettes where the r.h.s. is evaluated.
The Dirac strings are either forming closed loops or they are connecting
monopoles with anti-monopoles, $\delta \dual p[j] = \dual j$.
Whereas $\dual j$ is gauge invariant, the 2-form  $p[j]$ is not.

The simplest quantity describing the behavior of the monopoles is the
average monopole density
\be
\rho_{\rm mon} = \frac{1}{L^3} \sum_c |j_c| \,,
\ee
where $j_c$ is
the integer valued monopole charge inside the cube $c$ defined
in Eq.~\eq{jc}.

\begin{figure*}[!htb]
\begin{center}
\begin{tabular}{cc}
\includegraphics[scale=0.35,clip=true]{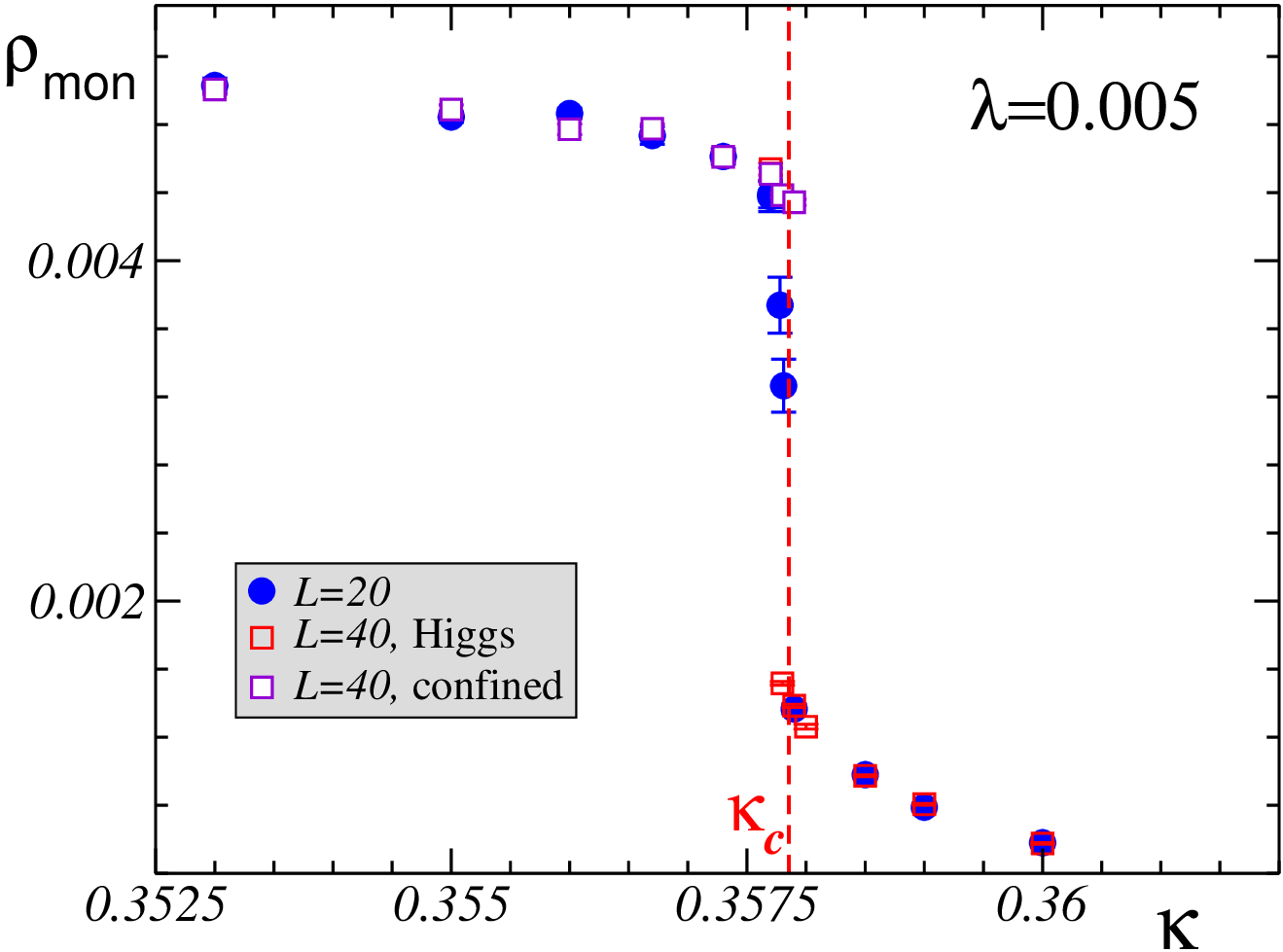} &
\includegraphics[scale=0.35,clip=true]{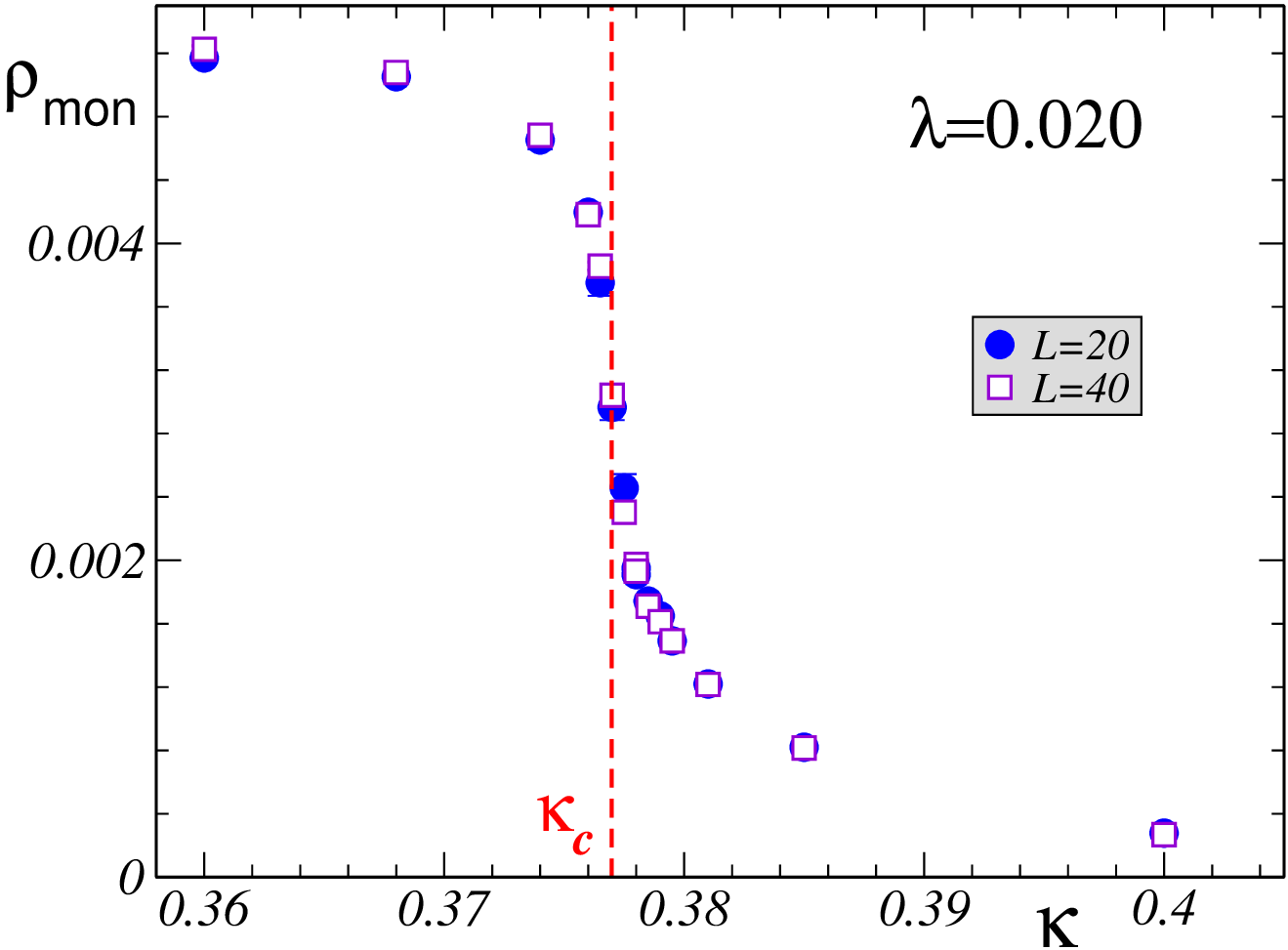} \\
  (a) & \hspace{5mm} (b)
\end{tabular}
\end{center}
\caption{Monopole density $\rho_{\rm mon}$
         near $\kappa_c$ for $20^3$ and $40^3$ lattice
         at $\lambda=0.005$ (a) and $\lambda=0.02$ (b).}
\label{fig:2}
\end{figure*}

The measured value of the monopole density at smaller and larger $\lambda$
are shown in Fig.~\ref{fig:2} for two lattice volumes $20^3$ and $40^3$ as
a function of $\kappa$.
The vertical lines are drawn to mark the pseudocritical $\kappa_c$.
Clearly the monopole density jumps from a nearly constant, finite
value to a small value crossing the transition from the confined to the Higgs
phase. At larger lattices a metastability region
is observed (in the form of a small hysteresis cycle) continuing the confined
and Higgs phase, respectively, into the critical region. For the lattice
$40^3$ the two metastable states at $\kappa_c$ are selected by the starting
conditions of the Monte Carlo run.
At higher $\lambda$ we observe a rapid but smooth decrease vs.
$\kappa$ near $\kappa_c$ precisely as in the London limit.

Finite size effects on the monopole density are very small, therefore, also
the susceptibility of the monopole density is practically unchanged with
increasing lattice size. As we have found earlier, the remaining non-vanishing 
density above the transition is due to dipole pairs forming a dilute dipole gas.
We will keep to these two values of the quartic self coupling
in order to study the influence of the different transition regimes on
the photon propagator.

\begin{figure*}[!htb]
\begin{center}
\begin{tabular}{cc}
\includegraphics[scale=0.35,clip=true]{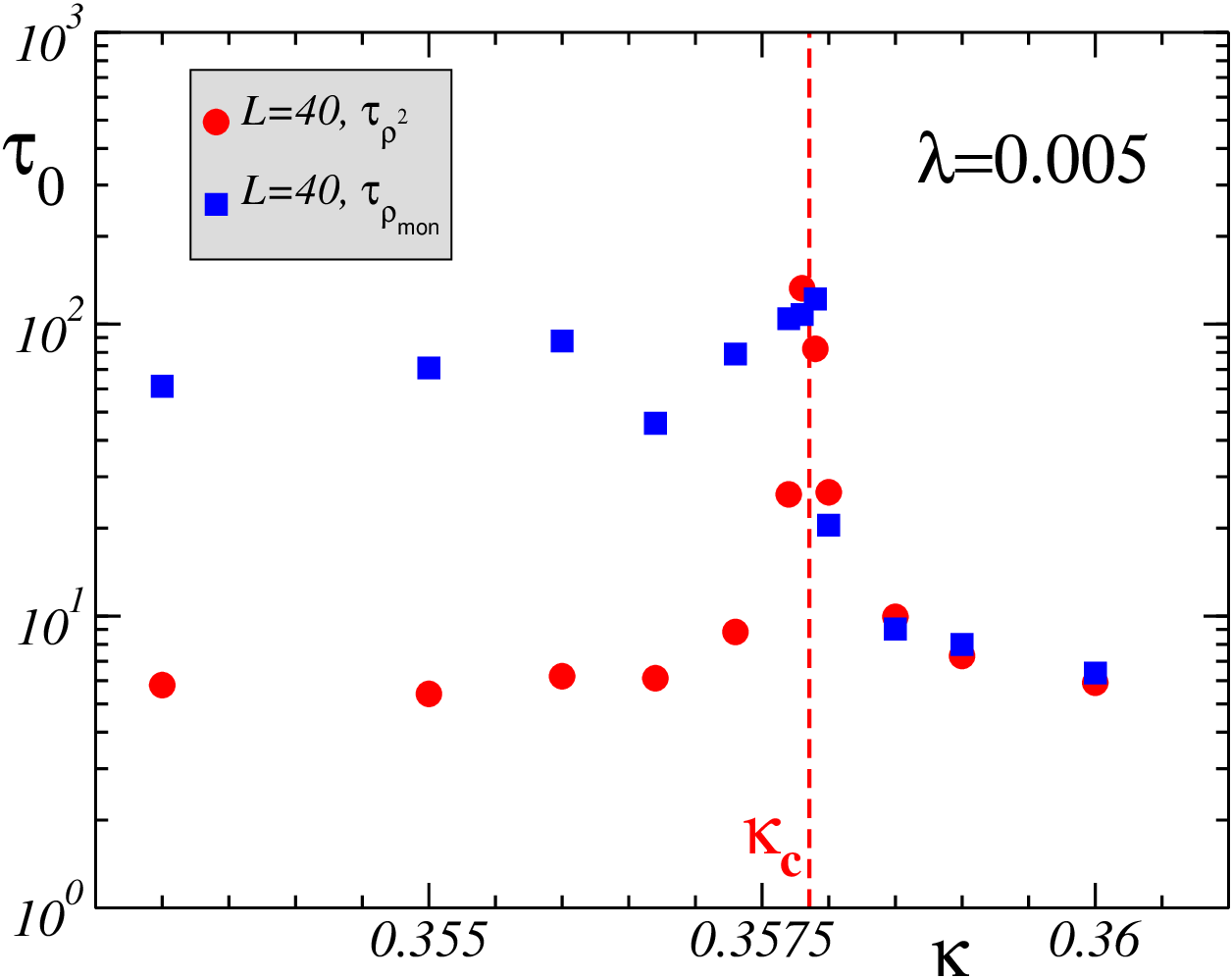} &
\includegraphics[scale=0.35,clip=true]{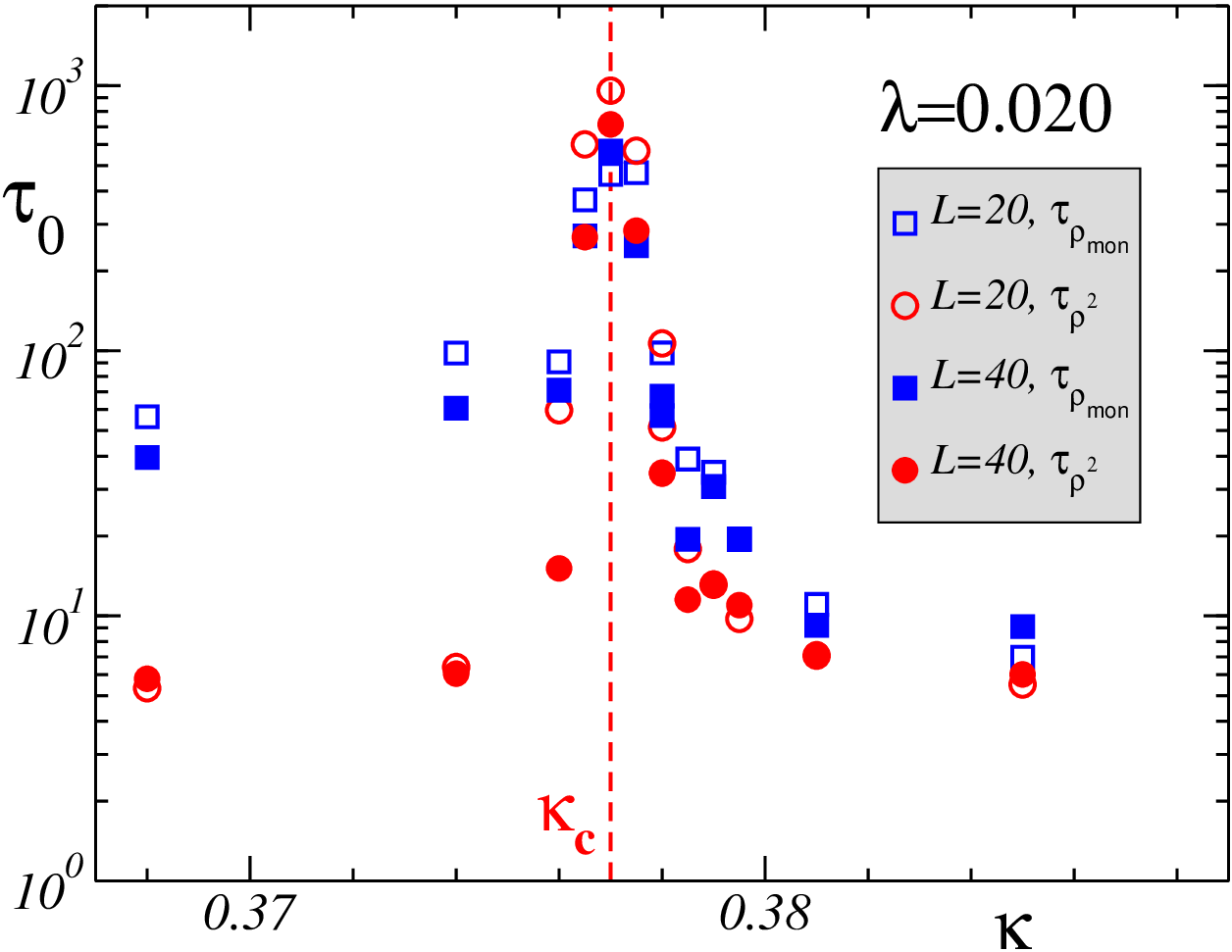} \\
  (a) & \hspace{5mm} (b)
\end{tabular}
\end{center}
\caption{The integrated autocorrelation times $\tau_O$  of $\rho^2$
 (circles) and $\rho_{\rm mon}$ (squares),
(a) in the first order region, (b) in the continuous transition region.}
\label{fig:autoCorr}
\end{figure*}

Finally we show in Fig.~\ref{fig:autoCorr}
the integrated autocorrelation time $\tau_O$ (for $\rho^2$ and $\rho_{\rm mon}$)
as function of $\kappa$. We recall that the autocorrelation time of an
observable $O$ is defined as
\be
   \tau_O = \frac{1}{2} +\sum_{\tau=1}^N C_O(\tau) \,, \quad  N\gg 1
\ee
in terms of the autocorrelation function $C_O(\tau)$.
In general, $C_O(\tau)$ is defined as
\be
C_O(\tau)= \frac{\langle O_t O_{t+\tau} \rangle - \langle O_t\rangle^2}
                {\langle O_t^2 \rangle - \langle O_t\rangle^2}
\ee
where the observable $O$ as measured at Monte Carlo ``time'' $t$
is folded with $O$ measured at $t+\tau$.
On the confinement side the autocorrelation time for the monopole density is
much larger than for the Higgs modulus squared.
For both quantities $\tau_O$ reaches ${\mathcal{O}}(100)$ in the first order
regime and up to ${\mathcal{O}}(1000)$ at critical coupling at larger $\lambda$.
Since for the latter case the autocorrelation
times of both $L=20$ and $L=40$ cubic lattices are roughly the same,
we have not to worry about critical slowing-down.
These autocorrelation times have been used to select statistically (almost)
independent configurations for measurements of the photon propagator
in the respective cases of the quartic Higgs self couplings.

\section{Landau gauge fixing with overrelaxation revisited}
\label{sec:gauge}

\subsection{Landau gauge and Dirac strings}

In order to calculate the (gauge dependent) photon propagator directly,
a special gauge has to be chosen, in our case the minimal Landau
gauge\footnote{For more details on the Landau gauge see~\cite{Chernodub:2002gp} 
and references therein.}.
This gauge is defined by finding the global maximum of the gauge functional
\be
F=\frac{1}{N_{\rm links}}\sum_{x,\mu} \cos(\theta^G_{x,\mu}) 
                                      \rightarrow {\rm max} \,.
\label{def:Landau_gauge}
\ee

The gauge field and Higgs field angles have to been transformed into this gauge
within gauge transformations $G$ with elements $g_x=\exp (i \omega_x)$
acting as
\beqn
      \theta_{x,\mu}^G &=& \left(\omega_x +\theta_{x,\mu} - \omega_{x+\mu} \right)_{2 \pi}\,.
      \\
      \varphi_x^G &=& \left( \varphi_x + \omega_x \right)_{2 \pi}
\eeqn
until the condition~(\ref{def:Landau_gauge}) is satisfied. The notation
$\left( \dots \right)_{2 \pi}$ indicates that values are restricted to the interval
$[-\pi,\pi)$ (modulo $2\pi$).

Every gauge transformation can be decomposed into a sequence of local gauge
transformations on each lattice site which would increase the
sum~(\ref{def:Landau_gauge}) restricted to neighboring links.
The corresponding angle $\omega_x$ is found easily as
\be
\tan \omega _x=
\frac{\sum_{\mu}\left(\sin \theta_{x,\mu}-\sin\theta_{x-\hat\mu,\mu} \right)}
{\sum_{\mu}\left(\cos \theta_{x,\mu}+\cos\theta_{x-\hat\mu,\mu} \right)}\,.
\label{tanomega}
\ee

The efficiency of the full optimization procedure depends crucially on the
landscape of the gauge functional in the space of all gauge transformations.
We have applied the local gauge transformation in a checkerboard fashion
(based on a separation into odd and even sublattices). It turns out that
overrelaxation allows us to speed up the finding of a local maximum
with respect to full gauge transformations $G$.
This was realized multiplying the gauge angles $\omega_x$ by an overrelaxation
parameter $1<\eta <2$.
Note that the extreme case $\eta=2$ does not change the local gauge
functional at all because it corresponds to a microcanonical update with
respect to the gauge functional $F$.
Fastest convergence was obtained for $\eta$ values of about 1.8-1.9
in agreement with earlier findings~\cite{Chernodub:2002gp}.

This algorithm will in general not lead to the absolute (global) maximum of
the gauge functional $F$ (\ref{def:Landau_gauge}). Typically it will get stuck in
one of the local maxima, which are called Gribov copies of each other and of the
(unknown) true maximum. As long as this non-uniqueness influences the
non-gauge-invariant observable
of interest, this is the origin of the so-called Gribov problem. Practically,
it is partially cured by repeating the same gauge fixing procedure, applying
it to random gauge copies of the original Monte Carlo configuration. The local
maximum with the largest value of $F$ is taken as the tentative global maximum.
This rests on the assumption that one of these random gauge copies might be
situated in the basin of attraction of the global maximum.

The number of gauge equivalent configurations produced to restart
the gauge fixing is denoted as $N_G$, and the iterative gauge
fixing generically leads to really different maxima. We have then
to content with the best out of all $N_G+1$ local maxima of
the gauge functional, and the observable of interest (the photon propagator) 
will depend on $N_G$.

For the purpose of the following discussion $\eta = 1.8$ has been chosen.
It will become clear that this choice is not advantageous.
A closer investigation of the gauge fixing procedure reveals a direct
relationship between the achieved value of the gauge functional and the
density of leftover (gauge dependent) Dirac strings~\footnote{For the
definition of the Dirac strings see Eq.~(\ref{Diracstrings}).
In four dimensional compact $U(1)$ the attention was drawn to Dirac sheets 
in~\cite{Bornyakov:1993yy}.  }
\be
   n_{\rm Dirac}= \frac{1}{N_{\rm links}}\sum_{^*P} |p_P[j]|
\ee
where the sum is taken over the links $^*P$ of the dual lattice. In particular,
maximizing the gauge functional leads to a minimization of the Dirac string
density as discussed in~\cite{Chernodub:2002gp}.
Note that this dependence is a result of the {\it local} gauge fixing procedure
which blindly acts with respect to the Dirac string content.

\begin{figure*}[!htb]
  \begin{tabular}{cc}
\includegraphics[scale=0.35,clip=true]{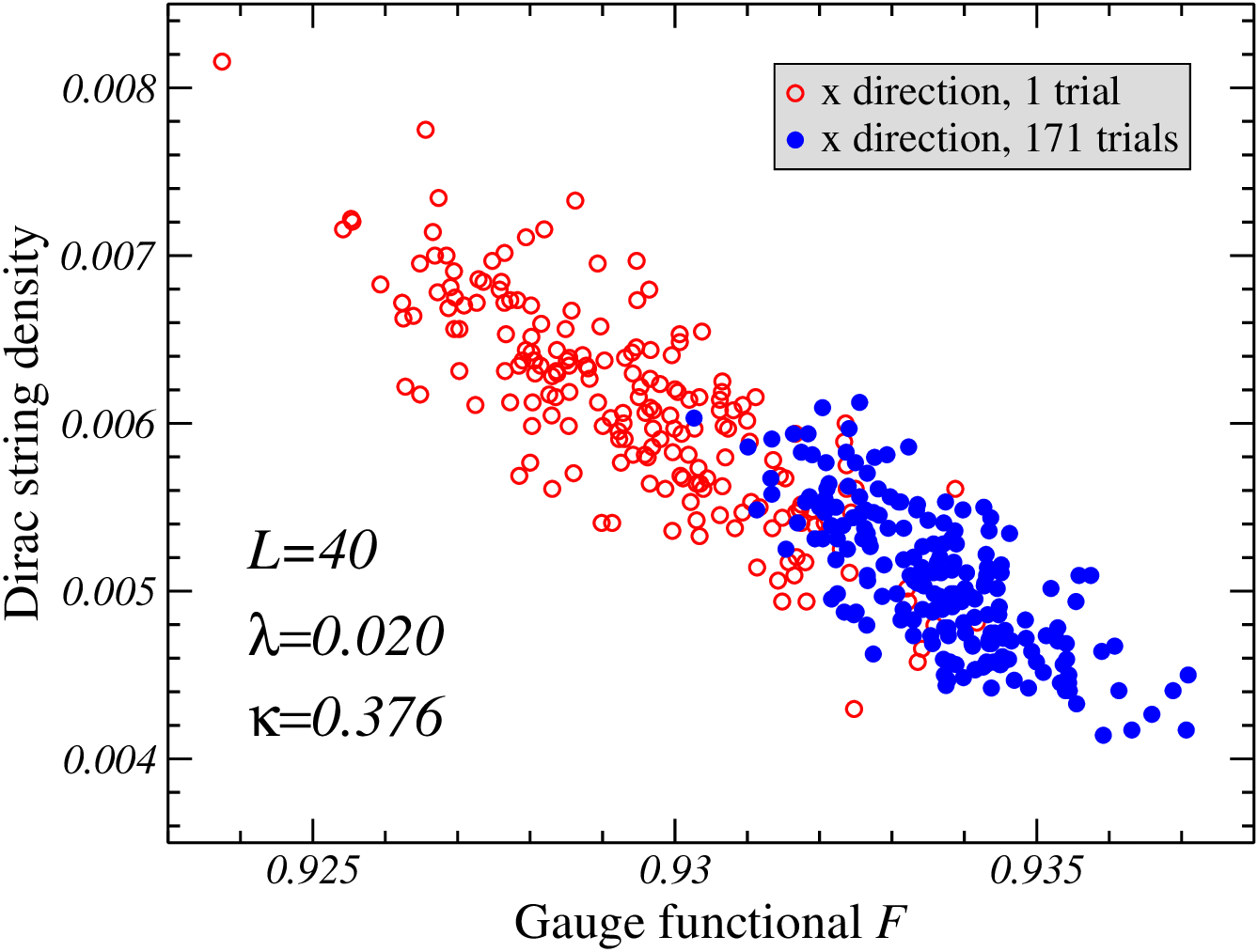} \hspace{5mm}&
\includegraphics[scale=0.35,clip=true]{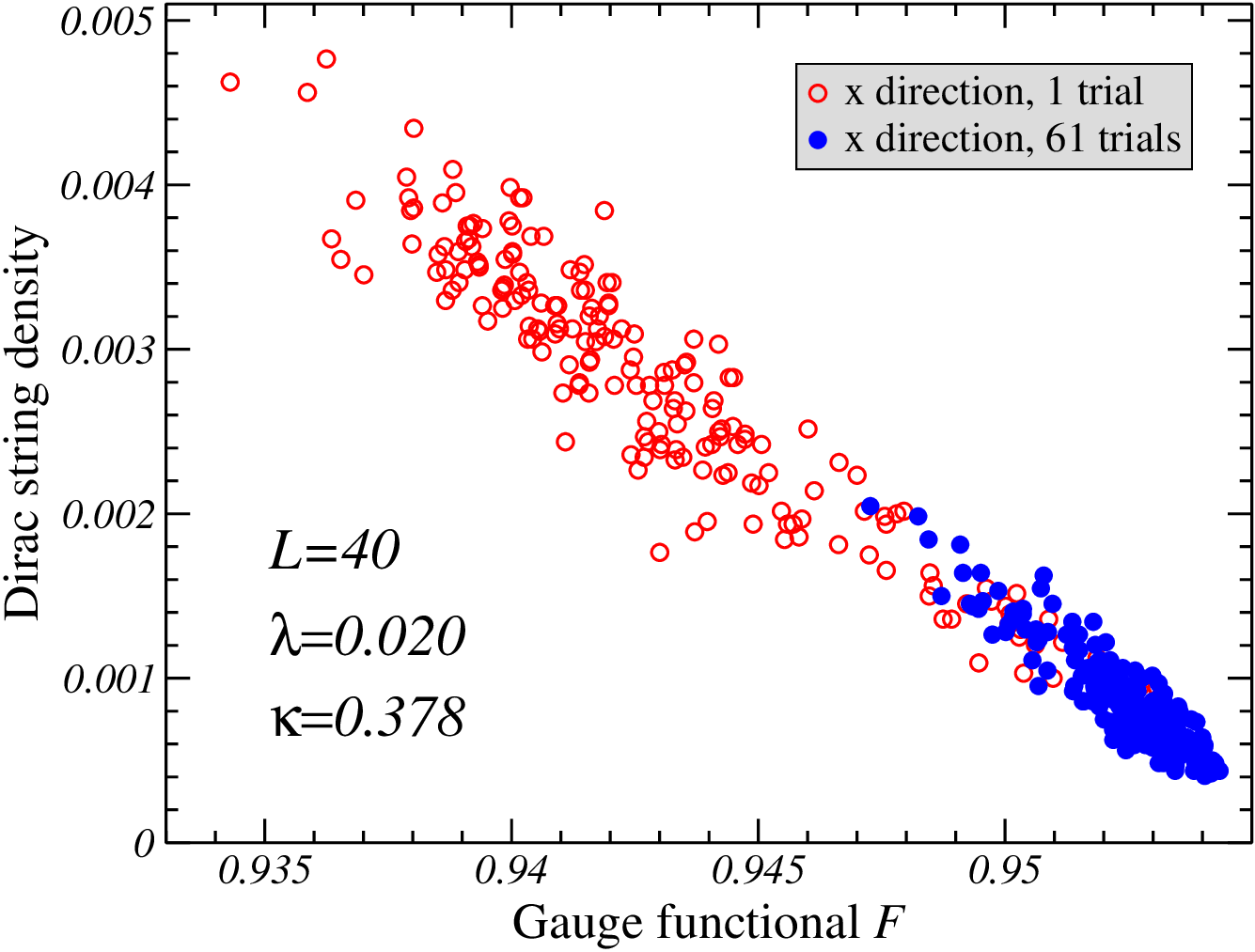} \\
  (a) & \hspace{5mm} (b)
  \end{tabular}
  \caption{(a) The highest gauge functional after 1 trial ($N_G=0$)
   and 171 trials ($N_G=170$) on the confined side of the continuous transition
   near $\kappa_c$;
  (b) the same with 1 and 61 trials, respectively, on the
   Higgs (deconfined) side.}
\label{fig:GvsdsD}
\end{figure*}

The different behavior in the confined and Higgs (deconfined) phase is
demonstrated in Fig.~\ref{fig:GvsdsD}
showing the Dirac string density (separately measured with respect to different
directions, but shown only for the $x$ direction) vs. the values of the gauge
functional for an ensemble of 195 independent configurations. Both quantities
have been finally achieved after inspecting the first or the best copy
among $N_G+1$ local maxima of the gauge functional $F$.

{}From the figures it is clear, that in both regions with a single attempt the
best realization of the gauge fixing has never been found.
In the Higgs region corresponding to deconfinement ($\kappa=0.378$), with
sufficiently many copies $N_G$ created for selection, in most cases the relative
maximum of the gauge functional is located in a small interval just below 0.955
with a Dirac string density below 0.001. Here one can hope finally to catch the
global maximum accompanied by a minimal density of Dirac strings with a modest
increase of the number of trials.

In the confining phase the behavior is different. Even with 170 copies under
inspection, the reduction of the Dirac string density and the increase of
the gauge functional is less efficient. Although the average of the gauge
functional is shifted towards higher values its spread compared to the original
spread remains rather large and the highest values of 0.938 are scarcely reached.
The relative decrease of the density of Dirac strings is poor.
We will see soon that such a far-from-perfect gauge fixing strongly influences
the propagator.

On the other hand, from the absolute values of $n_{\rm Dirac}$ it can be
concluded that Dirac lines wrapping around the torus are absent.
They cannot be blamed to spoil the photon propagator, an effect which otherwise
could be argued to be unphysical~\footnote{See Ref. \cite{Chernodub:2003bb} on
the photon propagator in the deconfined phase of cQED$_3$ where it was possible
do deal with this unphysical remnant of imperfect gauge fixing.}.
Essentially the same observations can be made in the case of the smaller
$\lambda$ value (in the neighborhood of the first order transition), but here
without the possibility to simply separate out the ''physical case''.
Thus, it is conceivable that a smarter {\it local} gauge fixing procedure
could be able to reduce the Dirac string density to a (gauge independent) minimum.

Before any gauge fixing is attempted, the gauge dependent Dirac strings
are either open lines (forming Dirac lines connecting monopoles and antimonopoles)
or closed loops. In the confined phase these strings form huge clusters where
the individual lines and loops cannot be resolved. It is impossible to uniquely
identify throughgoing lines/loops at points where more than two Dirac strings
meet. After some local maximum of the gauge functional has been found, the number
of closed loops has decreased considerably. Closed loops have the chance even to
be gauged away. Open strings which connect gauge-independent monopole-antimonopole
pair positions are only able to minimize their length (number of contributing
strings) during the gauge fixing procedure.

In an idealized picture with no leftover loops, a local maximum of $F$ is
associated with a realization of Dirac lines necessary to connect monopole pairs.
The global maximum then corresponds to a minimal realization. In the confined
phase we have a dense plasma of monopoles and antimonopoles with total charge
zero. Hence a lot of pairings are possible which corresponds to a huge number
of local maxima of the gauge functional. On the Higgs side (deconfinement)
only a small number of nearby (in lattice distances) monopole-antimonopole
pairs (a dilute dipole gas) is left. Consequently, the number of local maxima
is much lower than in the confined case. This picture explains why it is much more
unlikely to find the global maximum of $F$ in the confined phase compared to
the deconfined phase. The problem of the gauge fixing in the confined phase
belongs to the class of complex systems similar in complexity to the zero
temperature states in spin glasses.

\subsection{Choice of the algorithm}

The gauge fixing algorithm with continuous compact fields is very costly.
To speed up the algorithm several refinements have been done. The implementations
were tested for $\kappa<\kappa_c$ (confined side) because this is the most
critical region. Firstly, we tested the effect of restricting the link variables
to the discrete subgroup $\mathbb{Z}_N \subset U(1)$, {\it i.e.} mapping first
the gauge field variables to the closest $\mathbb{Z}_N$ realization.
Consequently, the gauge transformations were restricted to $\mathbb{Z}_N$, too.
We observed that approximating gauge field angles by 1 byte arrays ($N=256$) gave,
on the average, a lower gauge functional than using continuous compact fields.
Using 2 byte arrays ($N=65536$) the gauge functionals resulting from gauge fixing
using continuous angles and discrete angles, respectively, coincide within
statistical errors for some tested $\eta$ values.

Secondly, we used a preselection strategy. The $N_G+1$ gauge fixing attempts
can be performed at some chosen overrelaxation parameter $\eta\lesssim{}2$
with a rather weak stopping criterion, {\it i.e.} the lower limit for the change
of the gauge functional was still relatively large. The actual choice of the
best $\eta$ for the preselection stage is discussed below. Afterwards, among the
non-precisely gauge-fixed copies the trial with the highest gauge functional
was taken up again, and a final gauge fixing with a strong stopping criterion
was applied stepping back now to a smaller overrelaxation parameter $\eta=1.90$
leading to fastest final convergence during the final-gauge fixing stage.
The intuition behind this method is that during the preselection stage a
certain pairing of (anti-)monopoles is already chosen and the fine tuning
of the link angles is left to the second stage.

\begin{figure*}[!htb]
  \begin{tabular}{cc}
\includegraphics[scale=0.35,clip=true]{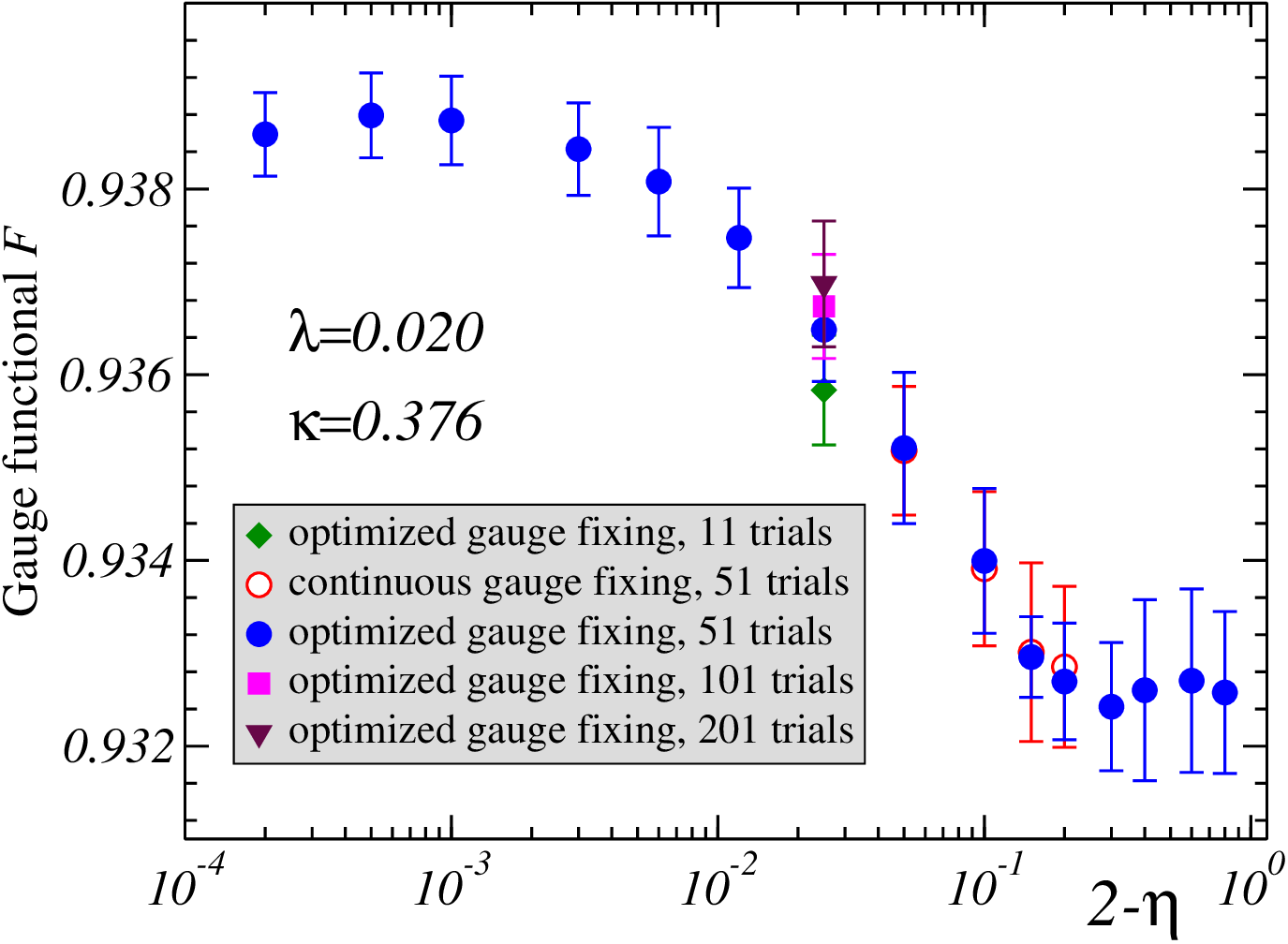} \hspace{5mm}&
\includegraphics[scale=0.35,clip=true]{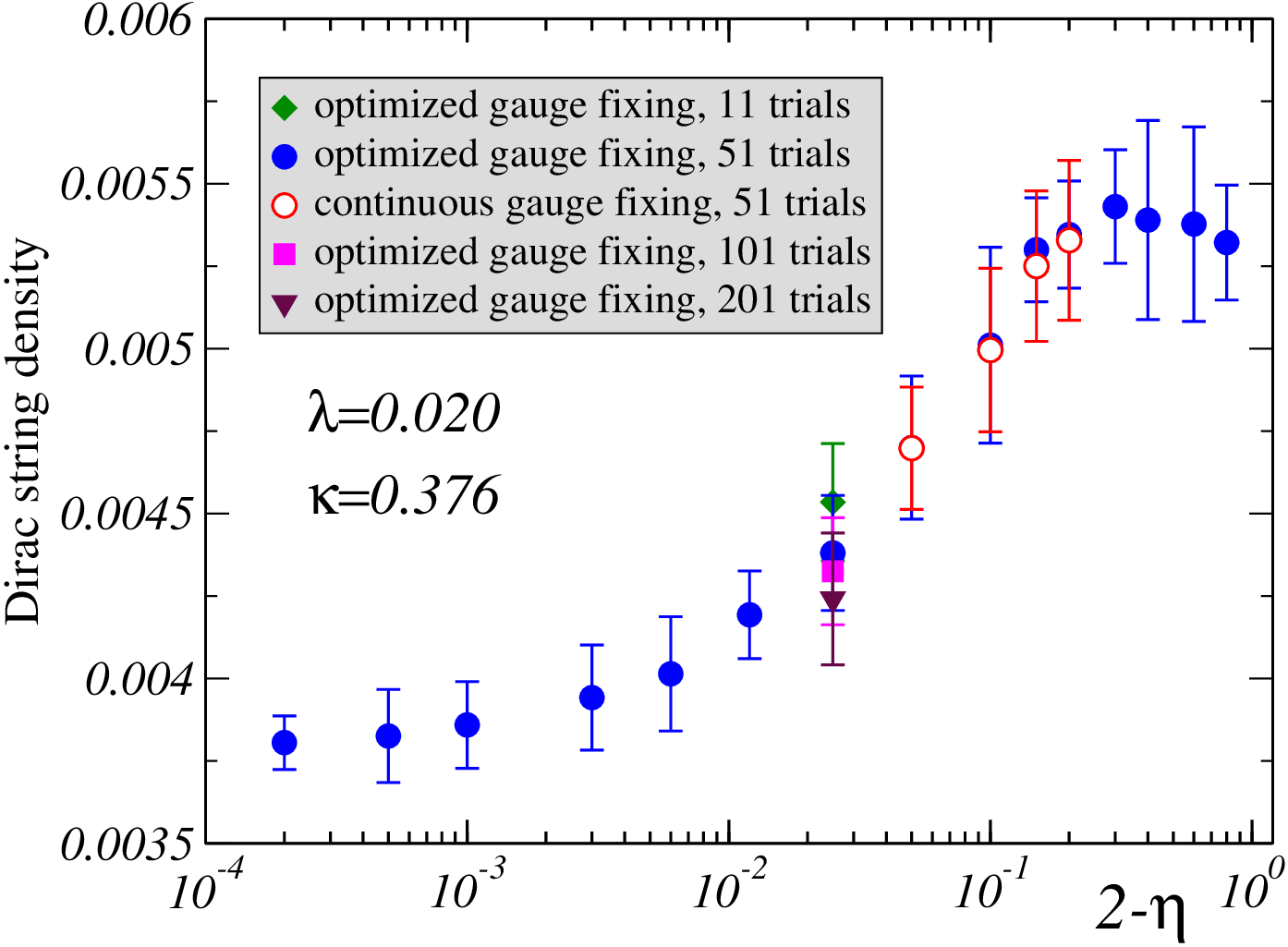} \\
  (a) & \hspace{5mm} (b)
    \end{tabular}
  \caption{
  (a) The average of the maximal gauge functional per
  configuration as function of $\eta$;
  (b) the same for the minimal Dirac string density.
  Continuous gauge fixing is selfexplaining.
  For details of the optimized gauge fixing see the text.}
\label{fig:EtaFdsD}
\end{figure*}

Both ideas have been then combined to use discrete gauge fixing in the
preselection stage and to apply continuum gauge fixing only to the best copy
obtained in the result of preselection.
For this purpose, a $\mathbb{Z}_N$ realization of the gauge field was used
first to construct a suitable $\mathbb{Z}_N$ gauge transformation, applying
the weak convergence criterion. Then, embedding the optimal discrete gauge
transformation into $U(1)$ as an initial guess and returning to the true $U(1)$
gauge field, the final continuous gauge fixing was done only for the most
promising copy.
The described combined procedure is called optimized gauge fixing
in Fig.~\ref{fig:EtaFdsD}.
Sensible stopping limits depend on the gauge functional landscape and thus on
the lattice size and the coupling parameters. They had to be chosen appropriately.
Using that optimized gauge fixing a speed-up of roughly a factor 10 has been found.

As mentioned above using $\eta=1.8$ might be not a good choice.
This is demonstrated in Fig.~\ref{fig:EtaFdsD}
showing the $\eta$ dependence of the gauge functional $F$ and of the Dirac
string density $n_{\rm Dirac}$ for $\eta$ values near (but below) two
using 20 independent measurements.
This dependence has been overlooked in our previous studies.
Only for the largest $\eta$ values (very close to $\eta=2$) we seem to reach
a new level of performance (higher gauge functional, lower Dirac string density).
In the figure for some selected $\eta$ values we also show measurements
comparing the results achieved with the optimized algorithm (using discrete
stage for preselection and preconditioning) with the standard algorithm using
only continuous gauge transformations.
As mentioned above, within statistics the results coincide.
At $2-\eta=0.02$ we show in addition how the change of $N_G$ influences
both gauge functional and Dirac string density.
In order to reach the same level by increasing the number of copies $N_G$
turned out to be much more expensive in computing time than increasing $\eta$.
Choosing $\eta$ close to 2 or, alternatively, applying several $\eta=2$
(the exactly microcanonical case) and $\eta<2$ overrelaxation steps in
alternating order, has led to a very similar convergence behavior.
In the first case the $\eta$ value was tuned and in the latter case we
have adjusted the number of microcanonical steps interspersed between
the overrelaxation steps in order to obtain the largest average of the
final gauge functional. The iteration number $N_{\rm iter}$
(iteration of local gauge transformations) needed in order to reach a
certain gauge functional has been found to be about the same in both cases.

Unfortunately, $N_{\rm iter}$ increases very fast if one tries to get at
higher values of the gauge functional using one of these methods.
Comparing $\eta=1.997$ with $\eta=1.80$ that number rises by a factor
of 15 to 20.
The highest value of $\eta$ which has led to an affordable $N_{\rm iter}$
in order to complete at least 50 independent measurements for the photon
propagator was actually $\eta= 1.997$ .
Using this value one reaches the edge of a new level of performance
both in the gauge functional
and the Dirac string density indicating that the effect of increasing $\eta$
further should become small.
Note that for $\eta$ close to 2 the algorithm behaves similar to a
simulated annealing technique.
Which of the techniques is computationally favorable, remains to be
studied. It has been observed that increasing $\eta$ is much more
efficient than increasing the number of trials $N_G+1$ using as measure
the computational workload, {\it i.e.} the total number of updates of the
local gauge transformations $(N_G+1) \times N_{\rm iter}$.

After these detailed studies we have chosen the following parameters
as a standard to perform the Landau gauge fixing in the measurements
of the gauge boson propagator:
As weak stopping criterion the change of the gauge functional
between two iterations had to become less than $1/(3 \times 40^3)$,
as strong stopping criterion we used $10^{-8}/( 3 \times 40^3)$.
The number of Gribov copies at $\eta=1.997$ was chosen to be 51.

\section{Results for the propagator near criticality}
\label{sec:results}

The photon propagator is the gauge-fixed ensemble average
of the following bilinear in $\tilde{A}$,
\beqn
D_{\mu\nu}({\vec p}) = \langle \tilde{A}_{ {\vec k},\mu}
                               \tilde{A}_{-{\vec k},\nu} \rangle \, ,
\label{def:propagator}
\eeqn
where $\tilde{A}$ is the Fourier transformed lattice gauge potential
(in lattice momentum space) related to the gauge links $\theta_{x,\mu}$
(fixed to minimal Landau gauge in coordinate space) via~\footnote{The
propagator in the continuum formulation is given by
$ (\beta/a) D_{\mu\nu}({\vec p})$.}
\beqn
\tilde{A}_{{\vec k},\mu} =
\left(\frac{1}{L^3}\right)^{1/2}
\sum\limits_x
\exp \Bigl( 2 \pi i~\sum_{\nu=1}^{3} \frac{k_{\nu}
(~x_{\nu}+\frac{1}{2}\delta_{\nu\mu}~) } {L_{\nu}} \Bigr)
~ \sin  \theta_{x,\mu} \,, \quad
k_\mu=0, \pm 1,..., \pm \frac{L_\mu}{2} \, ,
\label{def:fourier_transformation}
\eeqn
The vectors of lattice momenta ${\vec p}$ on the left hand side of
(\ref{def:propagator}) are related to the integer valued Fourier momenta
vector ${\vec k}=(k_1,k_2,k_3)$ as follows:
\beqn
p_\mu(k_\mu)=  \frac{2}{a} \sin \frac{ \pi k_\mu}{L_\mu}\,, \quad
\vec p =(p_1,p_2,p_3)
\,.
\label{def:momenta}
\eeqn
Thus the lattice equivalent of the same continuum $p^2$ can be realized
by different vectors $\vec k$ which eventually could reveal a breaking of
rotational invariance.

Assuming reality and rotational invariance
the most general tensor structure of the continuum propagator is
\beqn
D_{\mu\nu}({\vec p})=
P_{\mu\nu}({\vec p})~D(p^2) + \frac{p_\mu p_\nu}{p^2} \frac{F(p^2)}{p^2}
\label{def:tensor_structure}
\eeqn
with the three-dimensional  transverse projection operator
\beqn
P_{\mu\nu}({\vec p}) = \delta_{\mu\nu}- \frac{p_\mu p_\nu}{p^2} \, .
\label{def:transverse_proj}
\eeqn
The two scalar functions (form factors)  $D(p^2)$ and $F(p^2)$
can be extracted on the lattice from $D_{\mu\nu}({\vec p})$
by projection. Since $D_{\mu\nu}({\vec p})$ is only approximately rotationally
invariant, the form factors may be scattered rather than forming a smooth
function of $p^2$.
When the Landau gauge is exactly fulfilled $F(p^2) \equiv 0$.
On the lattice, this is actually the case as soon as one of the local maxima of
the gauge functional
(\ref{def:Landau_gauge})  is reached, with an accuracy which directly reflects
the precision at stopping of the gauge fixing iterations.
There is no possibility to monitor through $F(p^2)$ to what extent the global
maxima have been successfully found.

In order to measure the propagator, for lattice size $40^3$, we have considered
50 (almost) independent configurations (separated by 720 Monte Carlo sweeps)
on the confined/symmetric side and 100 configurations (separated by 360 Monte
Carlo sweeps) on the deconfined/Higgs side near $\kappa_c$ for
the two representative cases of $\lambda$ ($\lambda =0.005$ and $0.02$).
There were several reasons for choosing this lattice size. A large lattice
reduces the influences of the boundary and leads to smaller variances in
certain observables, e.g. the monopole density. Furthermore, in the first order
regime tunnelling was prohibited (mixed phases during measurement) and the
propagator properties of both phases in a metastability region could be
independently studied.
Finally, the relatively large size allows to measure the propagator
for more realizations of very small momenta where the propagator in
the confined phase is most sensitive.

\begin{figure*}[!htb]
\begin{center}
\begin{tabular}{cc}
\includegraphics[scale=0.35,clip=true]{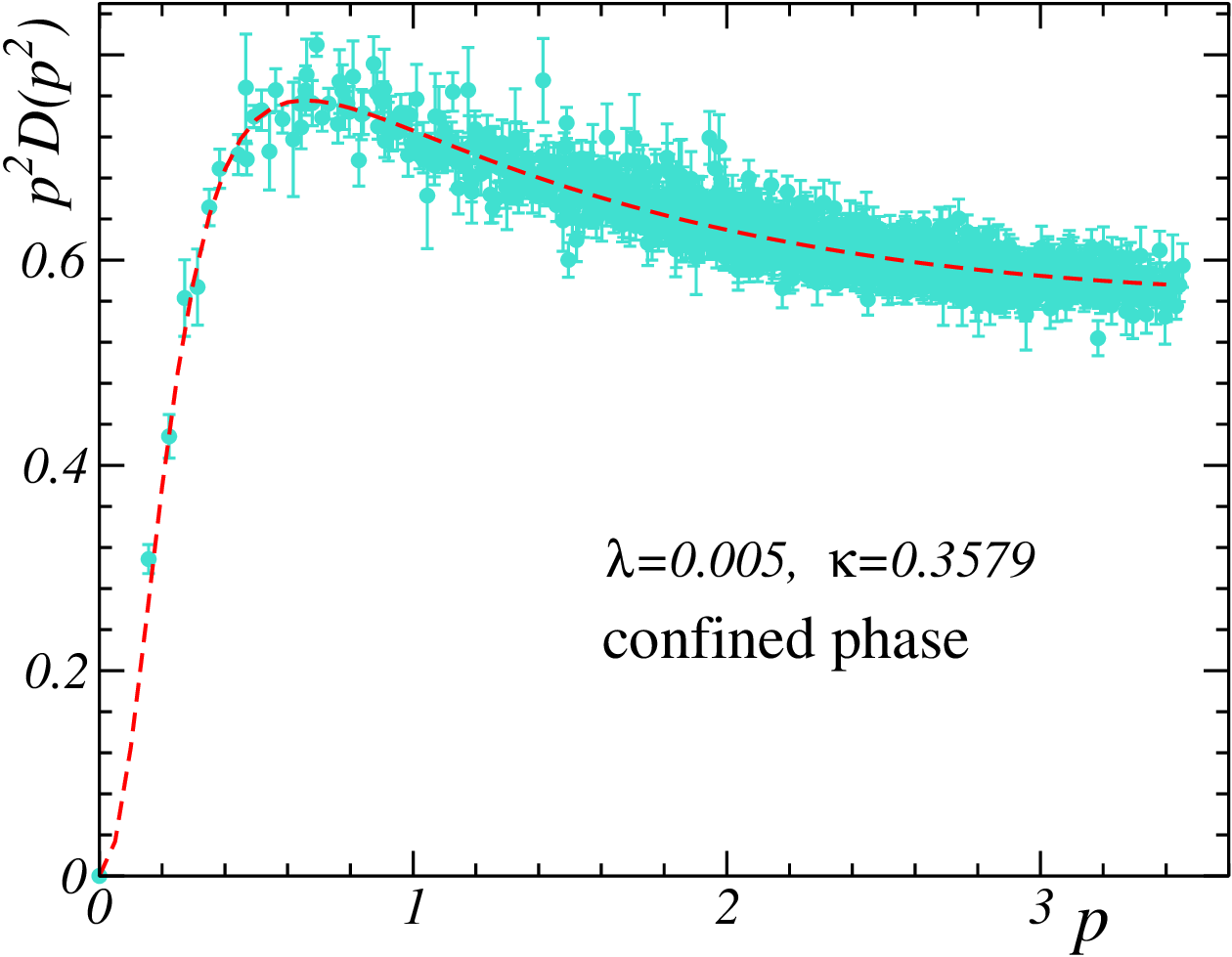} \hspace{5mm}&
\includegraphics[scale=0.35,clip=true]{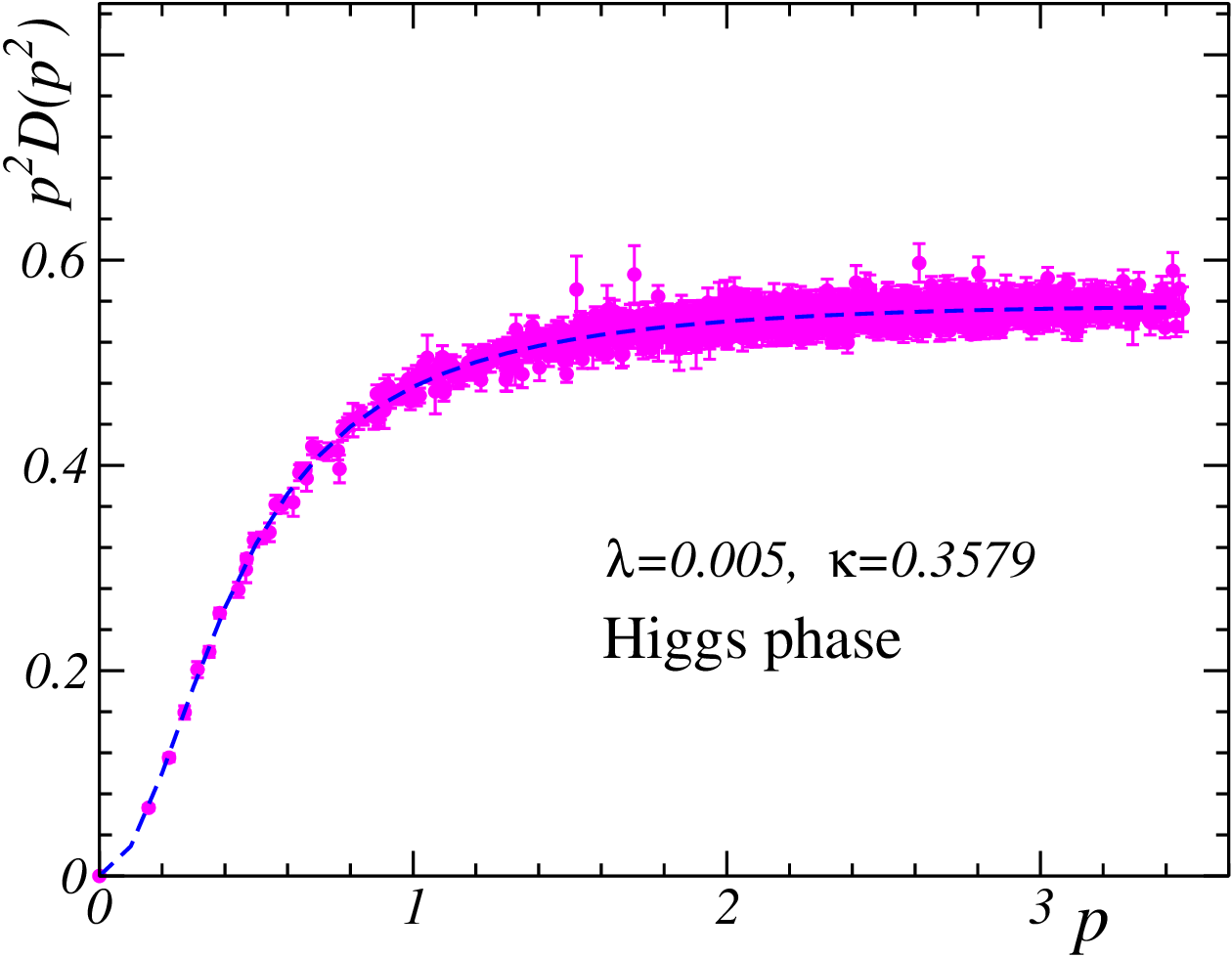} \\
  (a) & \hspace{5mm} (b)
\end{tabular}
\end{center}
\caption{Momentum dependence of the photon propagator form factor
  $D(p^2)$ multiplied by $p^2$
  for $\lambda=0.005$ at $\kappa_c=0.3579$
  together with the fit (\ref{eq:fit:total})
  in the confined/symmetric phase (a) and in the deconfined/Higgs phase (b).}
\label{propexamples}
\end{figure*}

In Fig~\ref{propexamples}
we present as example the form factor $D(p^2)$ multiplied by $p^2$
measured at the same $\kappa=\kappa_c$ in the metastability
region. The system was found either in the confined/symmetric and the
deconfined/Higgs phase depending on the initial conditions of the Monte Carlo
run.

To describe the scalar function $D(p^2)$ quantitatively
we applied the following fit ansatz
\beqn
  D(p^2) = \frac{Z\,m^{2 \alpha}}
  {\beta\left[p^{2 (1+\alpha)} + m^{2 (1+ \alpha)} \right]} + C\,,
  \label{eq:fit:total}
\eeqn
where $Z$, $\alpha$, $m$ and $C$ are the fitting parameters.
This fit has been successfully used to describe the propagators
of the finite and zero-temperature compact $U(1)$ gauge
model~\cite{CIS3,Chernodub:2002gp} in $2+1$ or $3$ dimensions, respectively
and the propagator in the London limit of the Abelian Higgs
model~\cite{Chernodub:2002en}.
The form is similar to some of Refs.~\cite{CurrentQCD} where the
propagator in gluodynamics has been studied. The meaning of the
fitting parameters in Eq.~\eq{eq:fit:total} is as follows: $Z$ is
the renormalization of the photon wavefunction, $\alpha$ is the
anomalous dimension, $m$ is a mass parameter. As shown in
Ref.~\cite{Chernodub:2002gp}, in cQED$_3$ this mass parameter coincides with
the Polyakov prediction~\cite{Polyakov} for the Debye mass,
generated by the monopole-antimonopole plasma.
The parameter $C$ corresponds to a $\delta$-like interaction in the coordinate
space and, consequently, is irrelevant for long-range physics.
We should remark that before fitting the data, the propagator values
have been averaged over all those lattice momenta realizing the same $p^2$.

The momentum dependence of the gauge boson
propagator is well described by the fitting function both in the
confined/symmetric and in the deconfined/Higgs phase.
The enhancement at intermediate momenta disappears (abruptly or
continuously) with increasing $\kappa$. 
In contrast, the suppression at very small $|{\vec p}|$
remains at all $\kappa$ values. We can conclude that the photon
propagator is always less singular in the infrared than a free one.
According to our fitting function~\eq{eq:fit:total} the propagator
is finite at $p^2=0$.

\begin{figure*}[!htb]
\begin{center}
\begin{tabular}{cc}
\includegraphics[scale=0.35,clip=true]{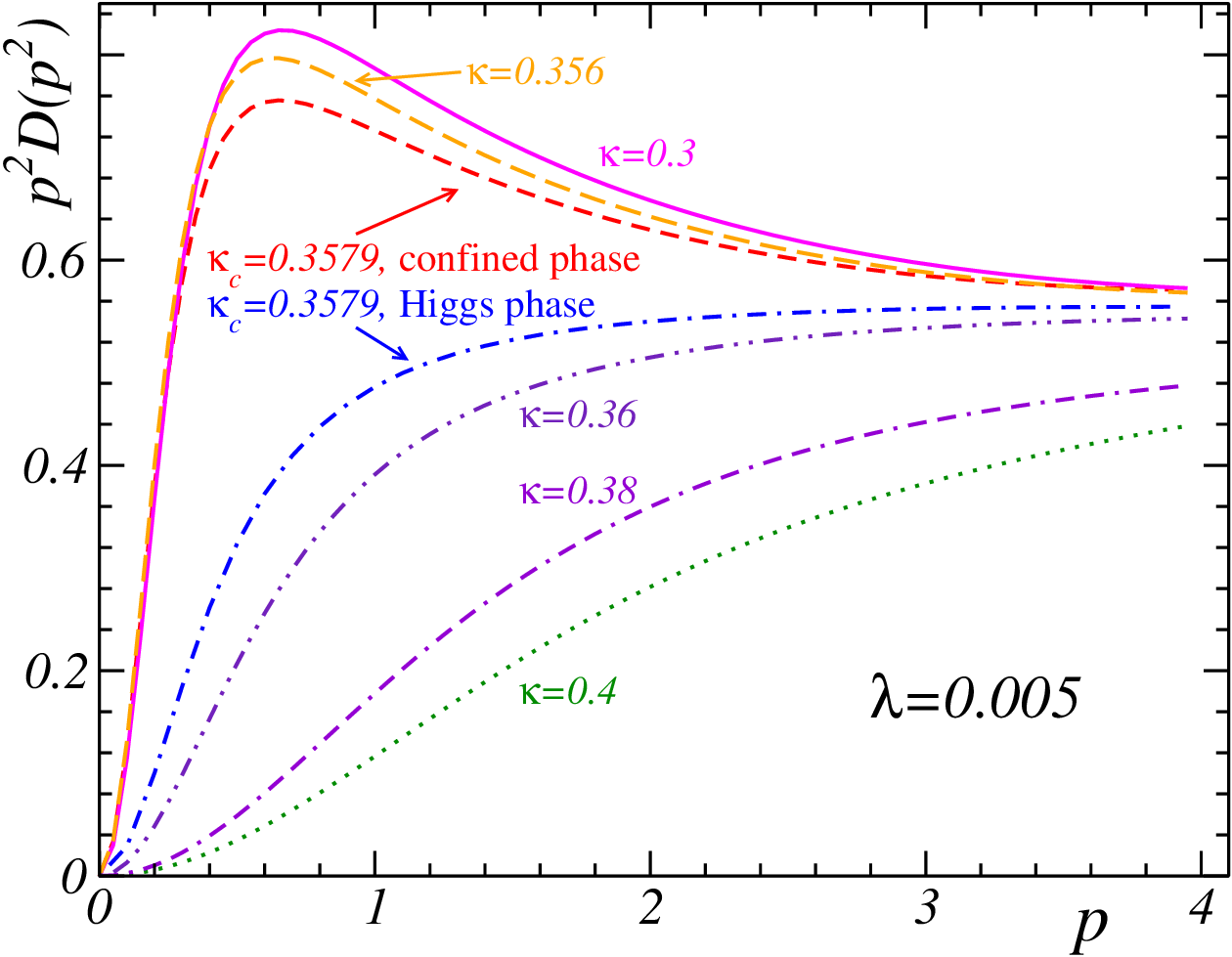} \hspace{5mm}&
\includegraphics[scale=0.35,clip=true]{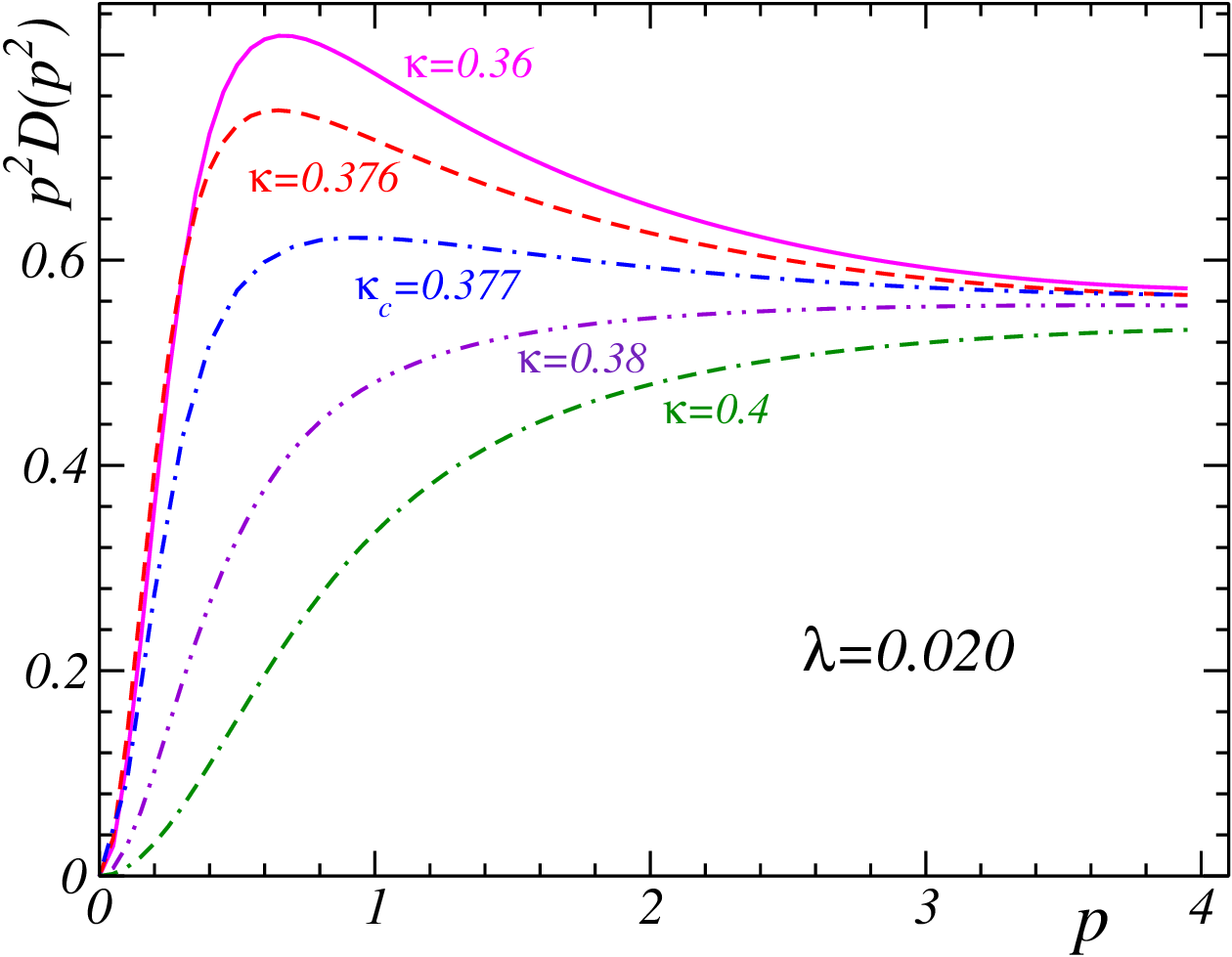} \\
  (a) & \hspace{5mm} (b)
\end{tabular}
\end{center}
\caption{Momentum dependence of the fitted photon propagator form factor
  $D(p^2)$ multiplied by $p^2$
  for different  $\kappa$'s near criticality
  for the first order (a) and continuous transition (b) regime.}
\label{propfits}
\end{figure*}

In Figs.~\ref{propfits}
we demonstrate how the momentum dependence of the propagator
changes when crossing the first order or continuous transition.
In the confinement region sufficiently away from the critical region
the propagator form factor is practically the same
irrespectively of the $\lambda$-value.
Approaching the critical $\kappa_c$ from below, the form factor $p^2~D(p^2)$
remains enhanced at intermediate momenta 
in the first order regime of small $\lambda$,
changing then abruptly to a free massive propagator in the Higgs phase
(deconfinement).
Contrary to that, $p^2 D(p^2)$ varies continuously with increasing $\kappa$
crossing the (continuous) transition at the higher $\lambda$ value.

\begin{figure*}[!htb]
  \begin{tabular}{cc}
\includegraphics[scale=0.35,clip=true]{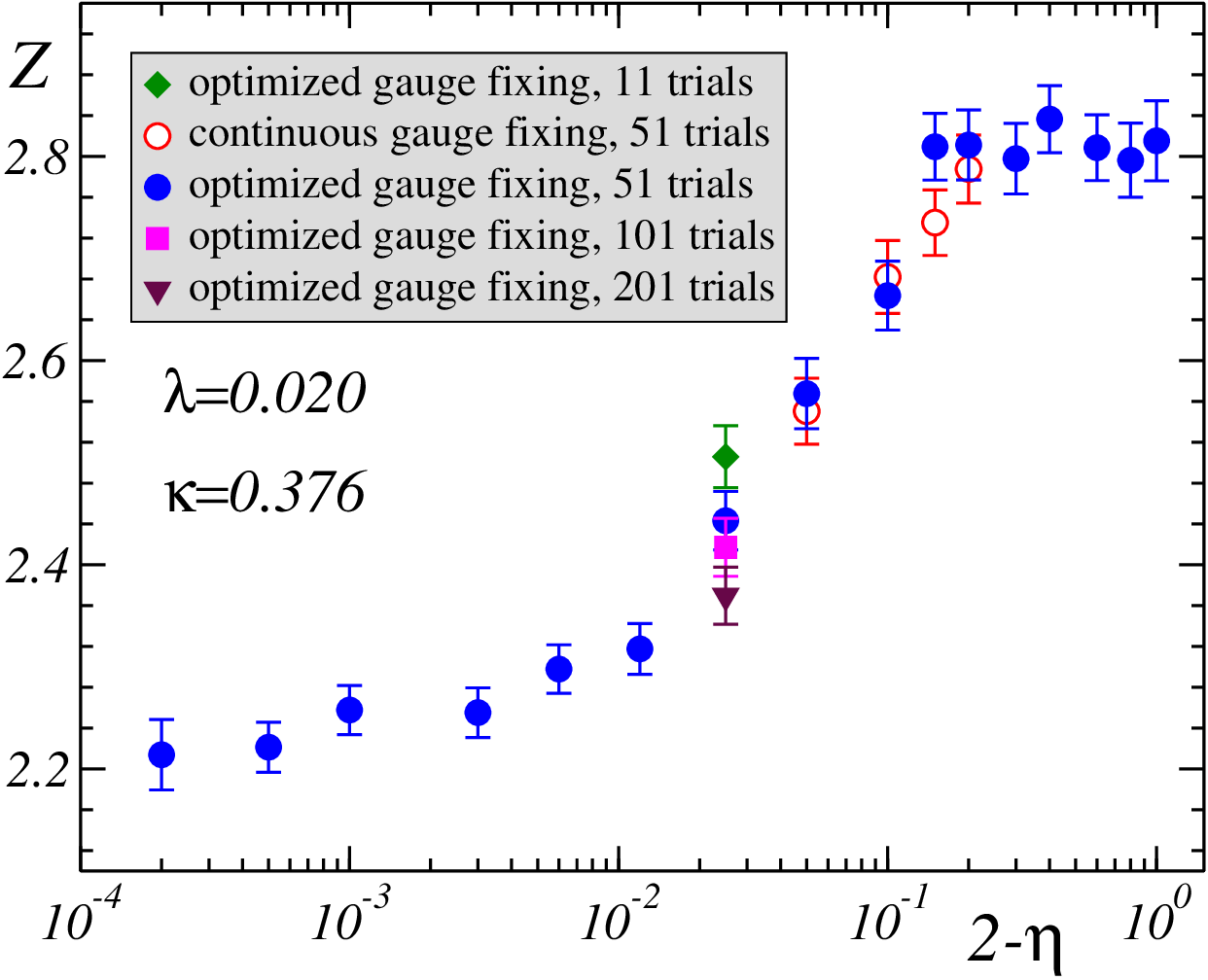} \hspace{5mm}&
\includegraphics[scale=0.35,clip=true]{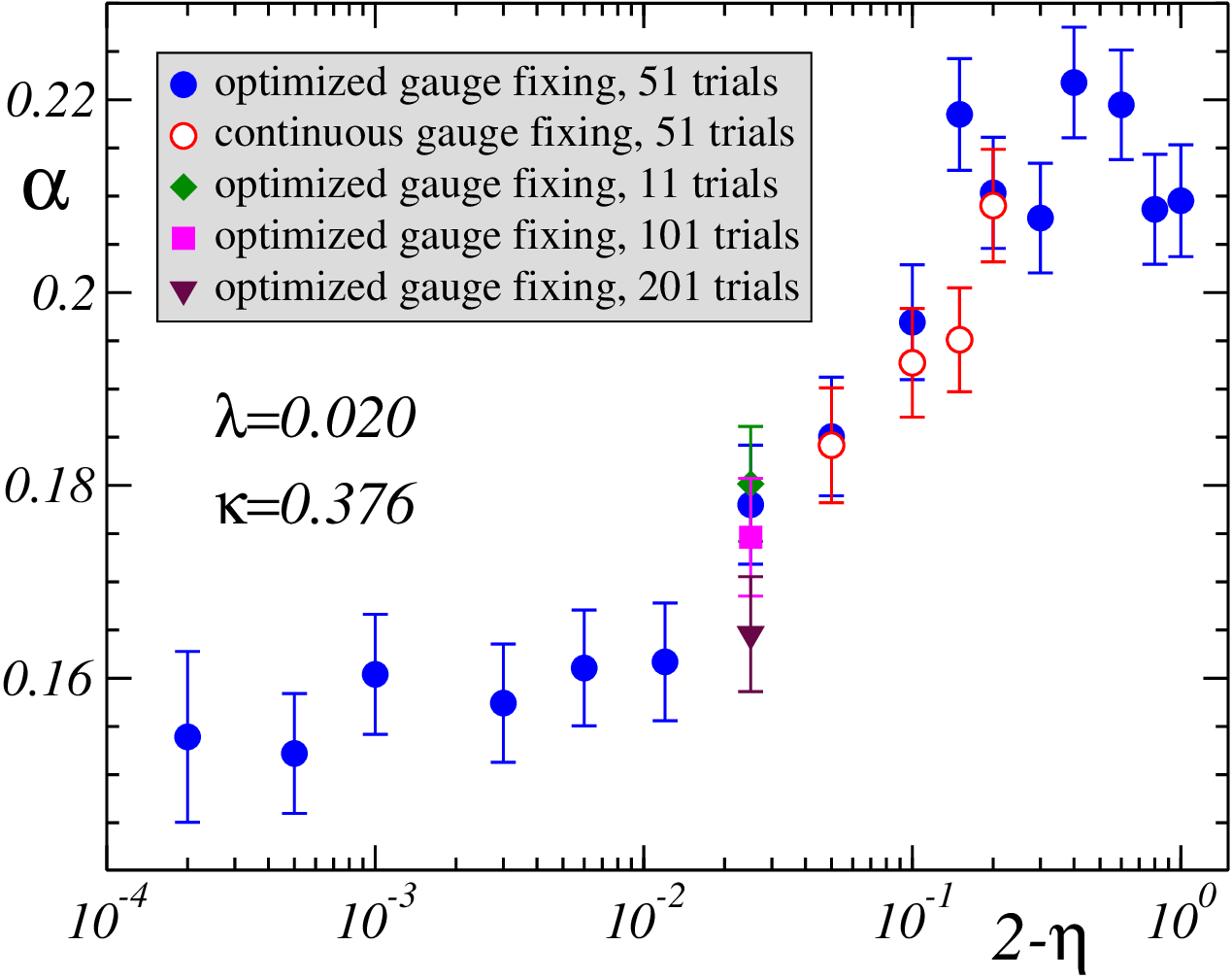} \\
    (a) & (b)
  \end{tabular}
  \caption{The fitted wave function normalization $Z$ (a)
   and anomalous dimension $\alpha$ (b) at $\lambda=0.02$
   and $\kappa=0.376$
   as functions of the overrelaxation parameter $\eta$.}
\label{fig:EtaZA}
\end{figure*}

In order to demonstrate the influence of a less efficient gauge fixing
procedure
on the shape of the emerging photon propagator we show in Fig.~\ref{fig:EtaZA}
the $\eta$ dependence of the fit parameters $Z$ and $\alpha$
on the confined side in the case of larger quartic Higgs self coupling
obtained in 20 measurements.
The dependence of the fit parameters resembles that of the gauge functional
and the Dirac string density presented in Fig.~\ref{fig:EtaFdsD} with a
transition from a less ''perfect'' level of performance
to a better one around
$2-\eta = 10^{-1} ... 10^{-2}$. This strongly suggests that these fit
parameters are closely related to the Dirac string density.
We have checked that the minor discrepancy between the fit parameters at 
$\eta=1.85$ using the optimized and continuous gauge fixing goes away with 
increasing statistics.

In contrast to this, the other fit parameters $m$ and $C$ are roughly
$\eta$ independent which points out that they depend only on the
(gauge independent) monopole content. Still, a minor $\eta$ dependence
of $Z$ and $\alpha$ might remain because we have used $\eta=1.997$
as a ''still affordable'' choice. Probably a different gauge fixing algorithm
has to be used to reduce this influence below the one-percent level.
Comparing the obtained fitting results with those obtained earlier,
with overrelaxation parameter $\eta=1.8$, for pure
$U(1)$~\cite{CIS3,Chernodub:2002gp} and for the Abelian Higgs model in the
London limit~\cite{Chernodub:2002en} we have to admit that the presented
numerical values for $Z$ and $\alpha$ in the confined phase are likely
to reach values smaller by roughly 20 to 25 per cent once a better tuned
$\eta$ is used. However, all qualitative results remain unchanged.

\begin{figure*}[!htb]
  \begin{center}
  \begin{tabular}{cc}
\includegraphics[scale=0.35,clip=true]{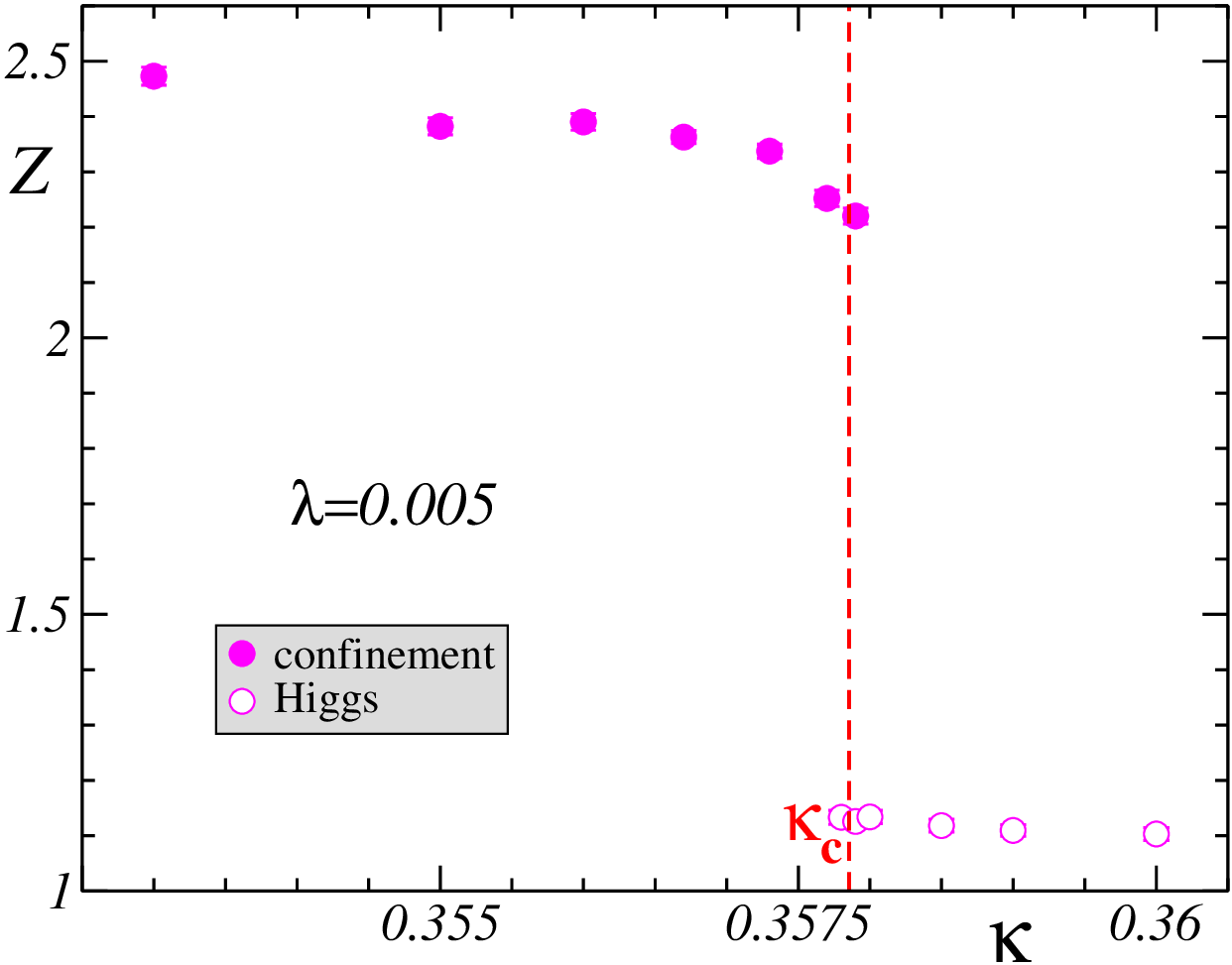} \hspace{5mm}&
\includegraphics[scale=0.35,clip=true]{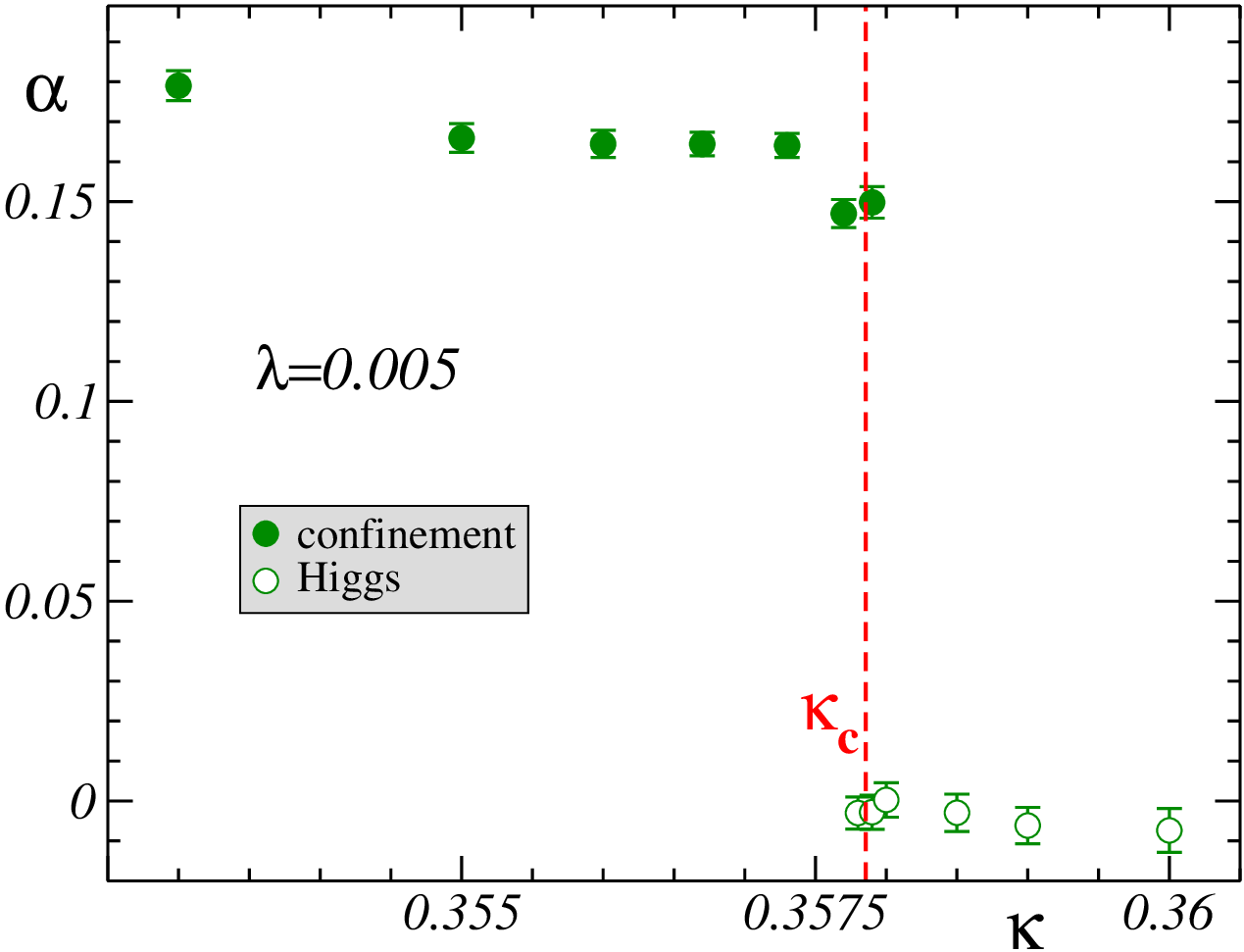} \\
    (a) & \hspace{5mm} (b)\\ \\
\includegraphics[scale=0.35,clip=true]{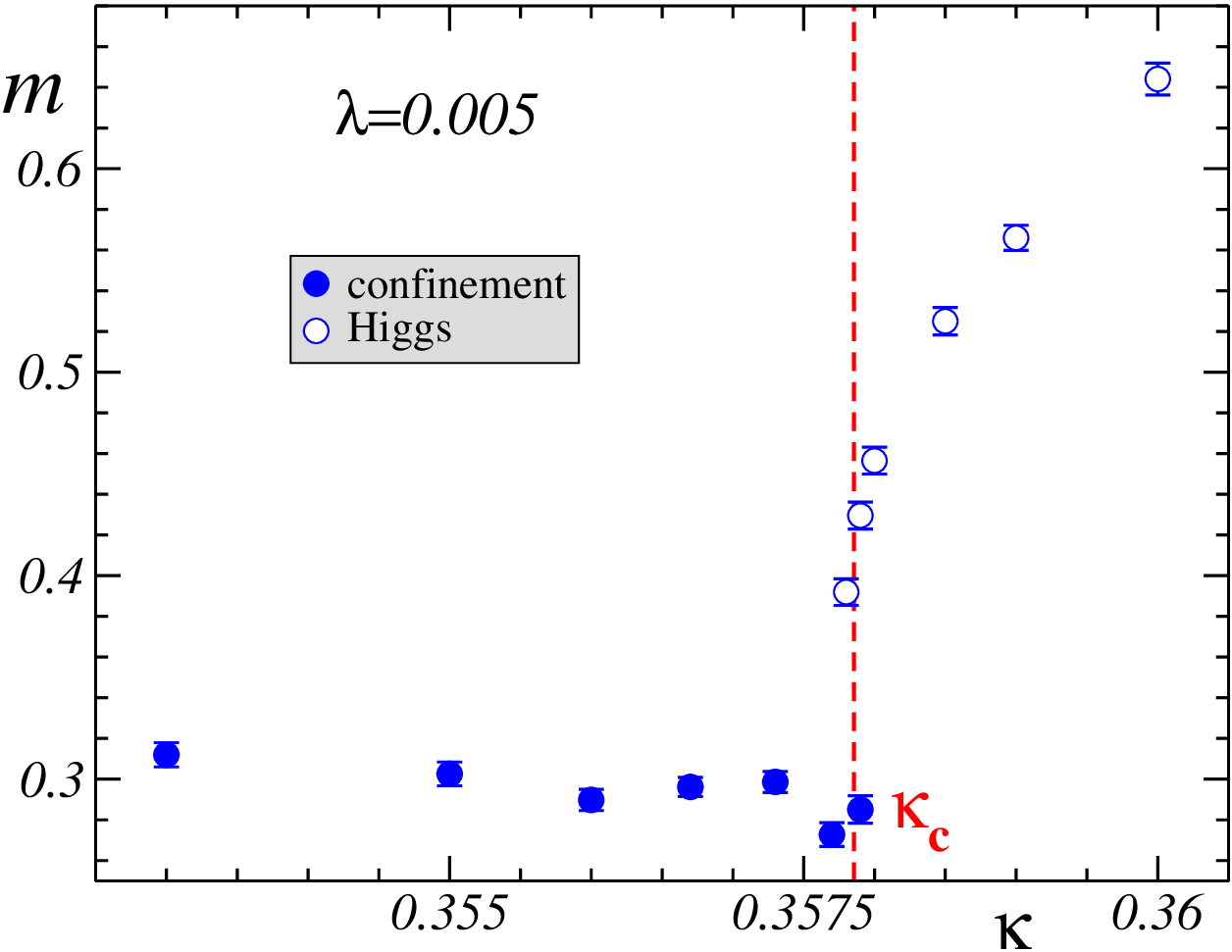} \hspace{5mm}&
\includegraphics[scale=0.35,clip=true]{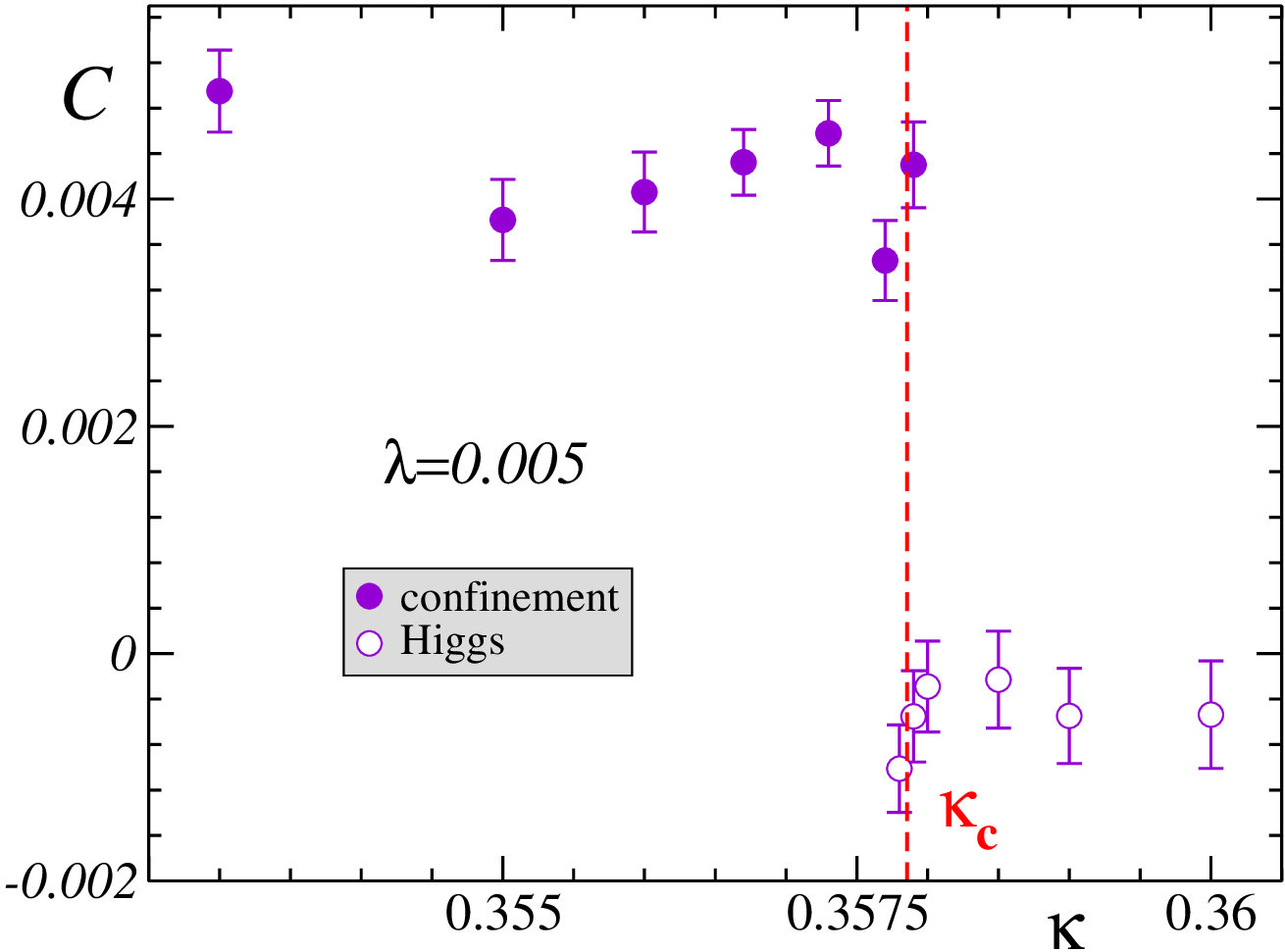} \\
  (c) & \hspace{5mm} (d)
  \end{tabular}
  \end{center}
  \caption{Fit parameters for the propagator at
  $\beta=2.0$  and $\lambda=0.005$
  (in the first order transition region)
   as function of $\kappa$:
  (a) $Z$;
  (b) $\alpha$;
  (c) $m$;
  (d) $C$.}
  \label{fig:firstorder}
\end{figure*}

The resulting fit parameters near the first order transition are presented in
Fig.~\ref{fig:firstorder} as functions of $\kappa$.
All parameters show a discontinuity at the phase transition. The metastability
can be seen clearly reaching the transition from below
(confined/symmetric phase) or from above (deconfined/Higgs phase).
The anomalous dimension jumps from a non-zero positive value at low $\kappa$
to zero in the Higgs phase.
The increasing mass of the photon propagator in the Higgs phase arises
from the Higgs mechanism, whereas in
the case of confinement the non-zero mass is due to the monopole plasma.
Note that the mass is only very weakly dependent on $\kappa$ up to $\kappa_c$
in agreement with the behavior of the monopole density
[compare Fig.~\ref{fig:2}(a)].

\begin{figure*}[!htb]
  \begin{center}
  \begin{tabular}{cc}
\includegraphics[scale=0.35,clip=true]{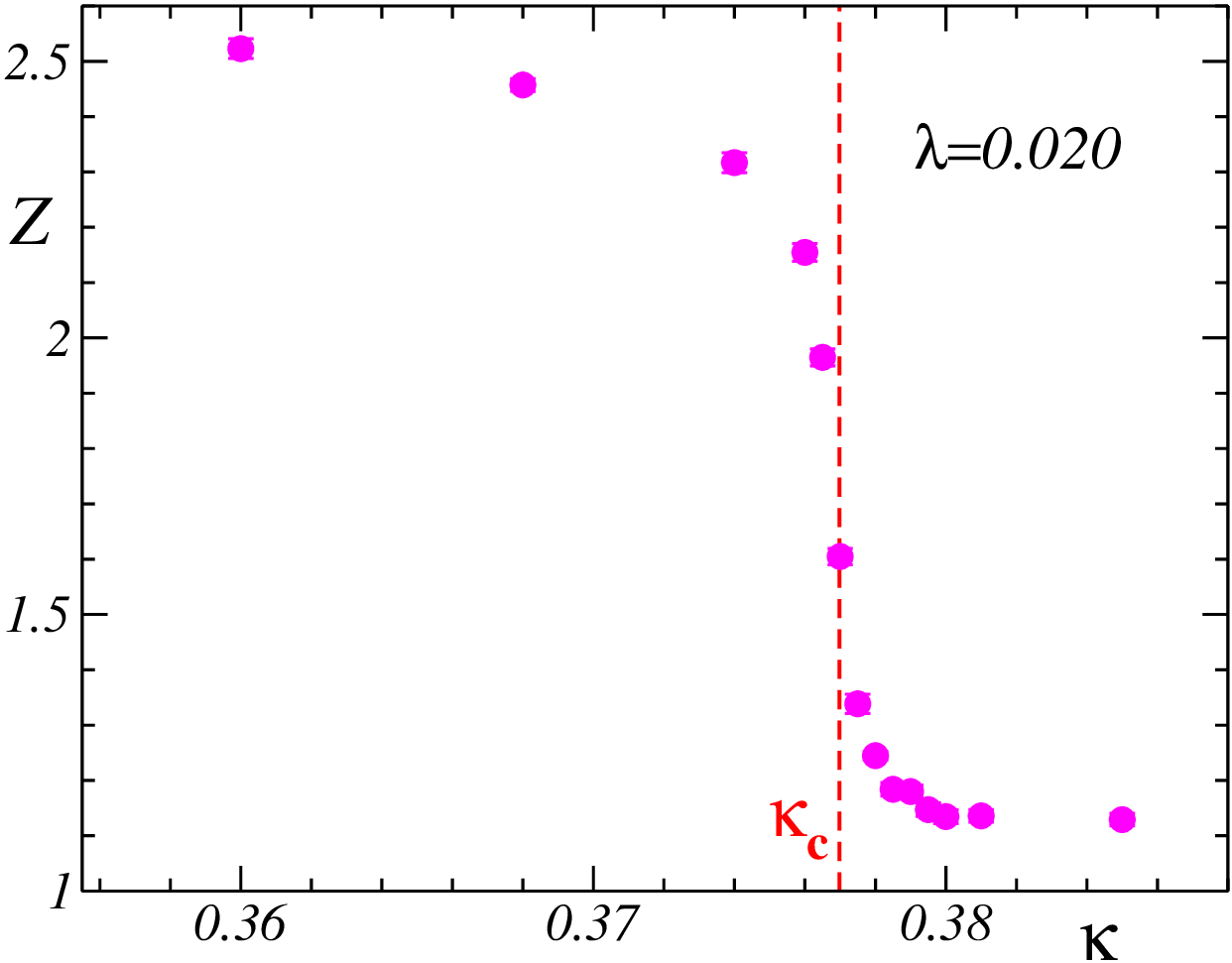} \hspace{5mm}&
\includegraphics[scale=0.35,clip=true]{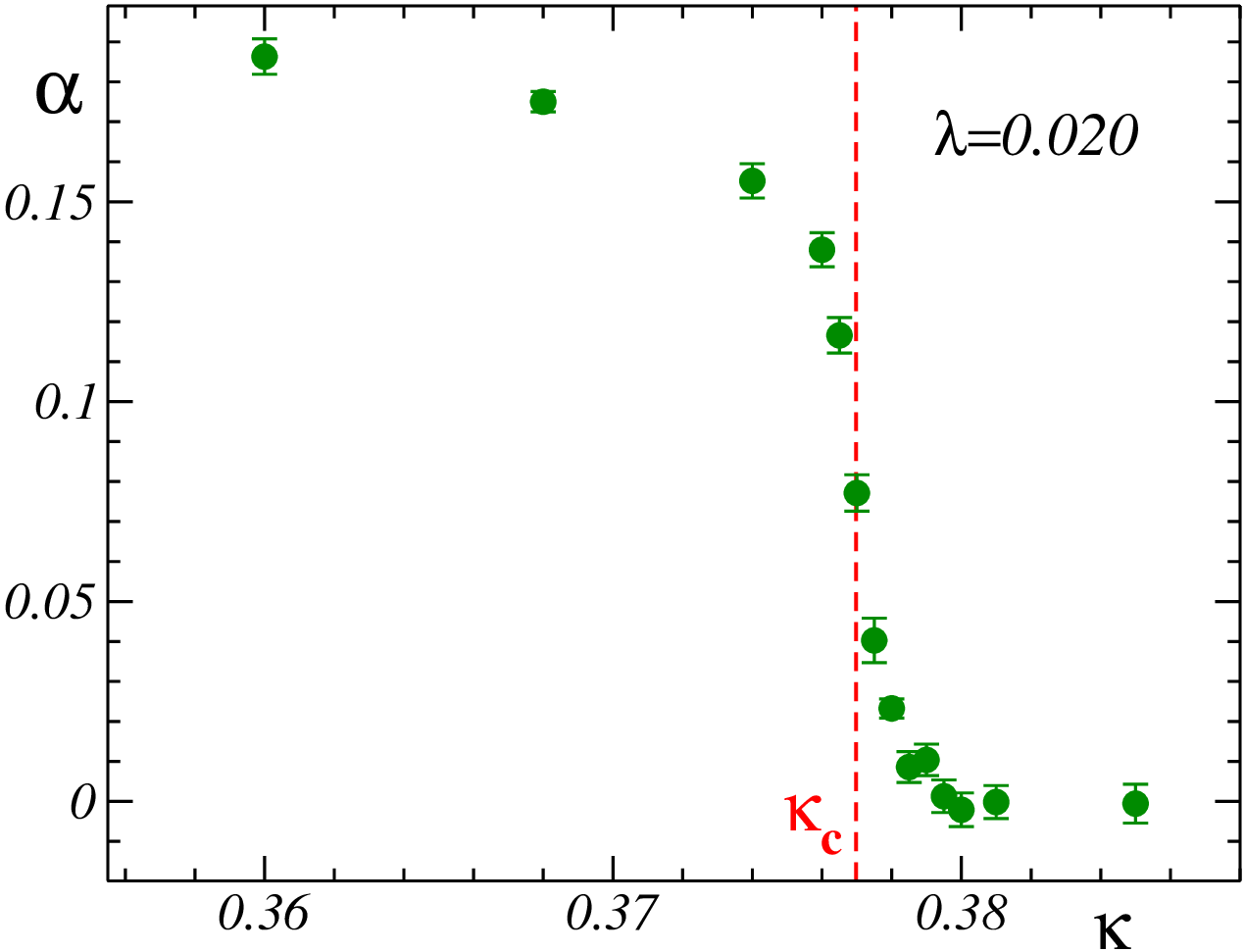} \\
    (a) & \hspace{5mm} (b)\\  \\
\includegraphics[scale=0.35,clip=true]{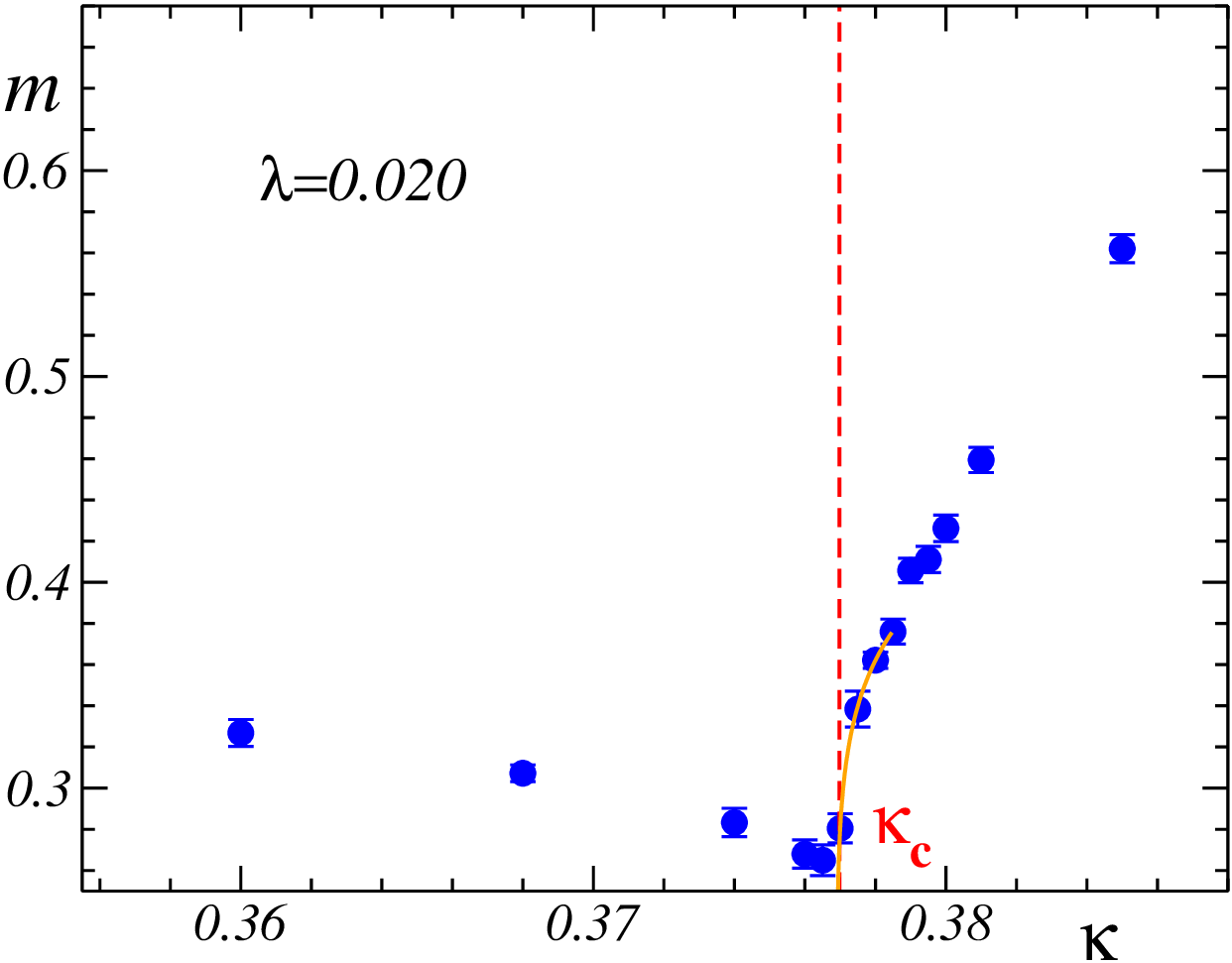} \hspace{5mm}&
\includegraphics[scale=0.35,clip=true]{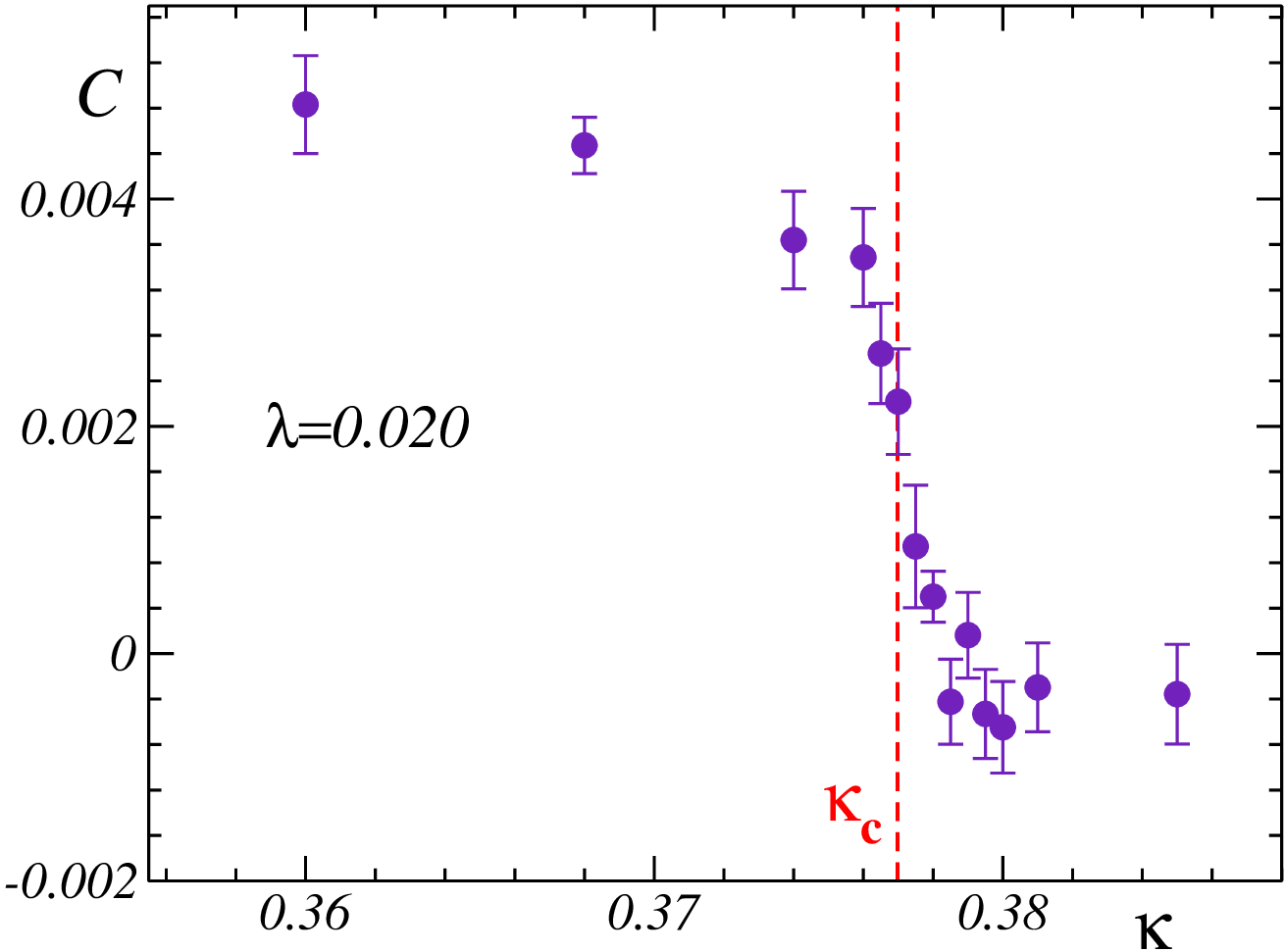} \\
  (c) & \hspace{5mm} (d)
  \end{tabular}
  \end{center}
  \caption{Fit parameters for the propagator at
  $\beta=2.0$  and $\lambda=0.02$
  (in the continuous transition region)
  as function of $\kappa$:
  (a) $Z$;
  (b) $\alpha$;
  (c) $m$;
  (d) $C$.
The fitted mass curve using Eq.~\eq{eq:mass:fit} is indicated by the 
solid line in Fig. (c).
  }
  \label{fig:nonfirstorder}
\end{figure*}

The corresponding results for the continuous transition regime are presented in
Fig.~\ref{fig:nonfirstorder}.
The fit parameters behave in complete analogy to the London limit:
In the region very close to $\kappa_c$,
the mass $m$ shows a minimum caused, as one
could guess, by the interference between perturbative mass and Debye
mass effects. Indeed, as $\kappa$ tends to $\kappa_c$,
the Debye mass gets smaller since
the density of the monopole-antimonopole plasma drops rapidly.
One the other hand, the perturbative mass term becomes more significant.
The interplay of these two tendencies results in the noticed minimum at
$\kappa \approx \kappa_c$.

We have fitted the behavior of the mass parameter, $m$, as function of $\kappa$
on the Higgs side near the transition  by the function
\beqn
m(\kappa) = \mu {(\kappa - \kappa^{\mathrm{fit}}_c)}^\nu\,,
\qquad \kappa > \kappa_c\,.
\label{eq:mass:fit}
\eeqn
The fit ansatz implies a vanishing mass at $\kappa_c$ what is actually not found
numerically. So we treat that ansatz only phenomenologically.
The best fit with $\chi^2/d.o.f. = 1.3$ allows to
determine $\kappa^{\mathrm{fit}}_c = 0.3767(3)$, which agrees with
the value from thermodynamic measurements given in Fig.~\ref{fig:1}(b) 
within errors. The estimated ``critical exponent'', $\nu=0.19(6)$, 
differs from the mean field exponent, $\nu_{MF}=1/2$
which had been confirmed 
in Ref.~\cite{ref:phase:prop1,ref:phase:prop2} for larger $\beta$.
The fit including the four nearest fitted masses is shown in 
Fig.~\ref{fig:nonfirstorder}(c) by the solid line.

{}From the results for the anomalous dimension in both regimes we conclude
that in cAHM with radial degrees of freedom the effect
of the monopole pairing on the propagator is the same as in cQED:
the anomalous dimension gets close to zero in the Higgs phase which is
dominated by the magnetic dipole gas. This change is continuous in
the continuous transition case (large $\lambda$), but discontinuous for the
first order phase transition (small $\lambda$).

Similarly to the anomalous dimension, the effect of the monopole
pairing on the renormalization parameter $Z$ of the photon
wavefunction in cAHM, shown in Fig.~\ref{fig:nonfirstorder}(a),
is remarkably similar to the cQED case observed in
Refs.~\cite{CIS3,Chernodub:2002gp}.
The wave function renormalisation $Z$ drops continuously
across the continuous transition,
but discontinuously at the first order transition
for $\kappa$ reaching  $\kappa_c$ from below.

\section{Conclusions}
\label{sec:conclusions}

We have carefully investigated the problems afflicting the Landau gauge 
fixing procedure based on overrelaxation.
An improved gauge fixing algorithm has been proposed and tested
which uses a finite subgroup in a preselection/preconditioning stage.
The computational gain in the expensive confinement region is a speed-up
factor around ten.
Such a stage could be useful for non-Abelian gauge groups as well.

We found that the gauge boson propagator of the
three-dimensional compact Abelian Higgs model with radially
active Higgs fields possesses a non-zero {\it positive} anomalous dimension
in the confinement phase (at small $\kappa$).
The momentum dependence of the propagator can be described by four
parameters: the mass, the anomalous dimension, the photon renormalization
function and the strength of the contact term.

At the first order phase transition (small Higgs self coupling) the parameters
of the propagator show a discontinuity while in the vicinity of the
continuous transition (large Higgs self coupling) the change of the parameters
is continuous. Apart from the difference caused by the nature of the phase
transition, the propagator behaves qualitatively similar to the propagator
calculated in the London limit. Thus we conclude that the main ingredient
which influences the gauge boson propagator is played by the compact phase
of the Higgs field.

Note that the anomalous dimension of the gauge field does not become
clearly negative neither in the confined phase nor in the deconfined
phase. As it is found in the case of the cQED~\cite{CIS3,Chernodub:2002gp}
the compactness of the gauge field provides a positive contribution to
the anomalous dimension while the role~\cite{Herbut,AnomalousMatter} of the
dynamical Higgs fields is to make the anomalous dimension negative.
Our study suggests that the positive contribution due to the compactness
of the gauge fields is stronger than the negative contribution from the
Higgs fields.

\acknowledgments{M.~N.~Ch. acknowledges a partial support from
the grants RFBR 01-02-17456, DFG 436 RUS 113/73910, RFBR-DFG
03-02-04016 and MK-4019.2004.2. E.-M.~I. is supported by DFG
through the DFG-Forschergruppe ''Lattice Hadron Phenomenology''
(FOR 465).}

\end{document}